\documentclass[12pt,nofootinbib]{article}
\usepackage[dvips]{graphicx}
\usepackage{amssymb}
\usepackage{amsmath}
\usepackage{epsfig}

\usepackage{epsf}
\usepackage{graphicx,epsfig}
\usepackage{amsfonts}
\usepackage{amssymb}

\def\del{\partial}
\def\CA{{\cal A}}

\def\k{K\"ahler}

\def\Mpl{M_{pl}}

\def\gev{{~{\rm GeV}}}

\makeatletter
\renewcommand\section{\@startsection {section}{1}{\z@}%
                                 {-3.5ex \@plus -1ex \@minus -.2ex}
                                   {2.3ex \@plus.2ex}%
                                   {\normalfont\large\bfseries}}
\renewcommand\subsection{\@startsection{subsection}{2}{\z@}%
                                   {-3.25ex\@plus -1ex \@minus -.2ex}%
                                     {1.5ex \@plus .2ex}%
                                     {\normalfont\bfseries}}
\renewcommand\subsubsection{\@startsection{subsubsection}{3}{\z@}%
                                   {-3.25ex\@plus -1ex \@minus -.2ex}%
                                     {1.5ex \@plus .2ex}%
                                     {\normalfont\itshape}}
\makeatother

\newcommand{\Letter}{
\setlength{\textwidth}{16.5cm}
   \setlength{\textheight}{22.6cm}
    \hoffset=-0.5in
\voffset=-2.1cm }

\Letter

\newcommand{\be}{\begin{equation}}
\newcommand{\ee}{\end{equation}}
\newcommand{\bea}{\begin{eqnarray}}
\newcommand{\eea}{\end{eqnarray}}
\newcommand{\barr}{\begin{array}}
\newcommand{\earr}{\end{array}}

\def\beq{\begin{equation}}
\def\eeq{\end{equation}}
\def\be{\begin{equation}}
\def\ee{\end{equation}}
\def\bea{\begin{eqnarray}}
\def\eea{\end{eqnarray}}

\DeclareRobustCommand{\SkipTocEntry}[4]{}

\begin{document}

\begin{titlepage}

\setcounter{page}{1} \baselineskip=15.5pt \thispagestyle{empty}

\begin{flushright}
\end{flushright}
\vfil

\begin{center}
{\LARGE String Cosmology: A Review}

\end{center}
\bigskip\

\begin{center}
{\large Liam McAllister$^{1}$ and Eva Silverstein$^{2}$}
\end{center}

\begin{center}
${}^1$\textit{Laboratory for Elementary Particle Physics, Cornell University, Ithaca NY 14853}
\vskip 3pt
${}^2$\textit{SLAC and Department of Physics, Stanford University, Stanford CA 94305}
\end{center} \vfil

\noindent We give an overview of the status of string cosmology.  We explain the motivation for the subject,
outline the main problems, and assess some of the proposed solutions.  Our focus is on those aspects of
cosmology that benefit from the structure of an ultraviolet-complete theory.

\vfil
\begin{flushleft}
\today
\end{flushleft}

\end{titlepage}

\newpage
\tableofcontents
\newpage

\section{Yet Another String Cosmology Review}

The past decade has seen tremendous advances in cosmology.  The discovery of dark energy \cite{Supernova}
crushed widely-held expectations that some unknown mechanism might set the cosmological constant to zero.  At
the same time, a flood of data on the cosmic microwave background (CMB) \cite{Observations} has revealed an
early universe in striking agreement with the basic predictions of inflationary scenarios.  The resulting
paradigm of a universe undergoing inflation \cite{Inflation} at early times, and dominated by cold dark matter
and dark energy at late times, has sometimes been referred to as a Standard Model for cosmology.

However, the situation is far more mysterious than the comparison with the Standard Model of particle physics
would suggest. Inflation provides a robust field-theoretic mechanism that addresses many cosmological problems
-- and that made predictions \cite{Mukhanov:2005sc} strikingly verified by CMB experiments  -- but the data does
not yet pin down anything close to a precise model.  Similarly, we lack a complete understanding of dark energy;
the best explanation to date of the dark energy scale is a selection effect \cite{Weinberg:1987dv}. Both of
these subjects, although modeled at a basic level within the framework of effective field theory, are closely
connected to challenging issues in quantum gravity.


At the same time, substantial theoretical progress in string theory has brought forth a diverse new generation
of cosmological models, some of which are subject to direct observational tests. One key advance is the
emergence of methods of moduli stabilization. Compactification of string theory from a total dimension $D$ down
to four dimensions introduces many gravitationally-coupled scalar fields -- moduli -- from the point of view of
the four-dimensional theory.  Runaway or light moduli are extremely problematic in cosmology, and in any
realistic model they must be metastabilized by an appropriate balance of forces.  Extensive work in the subject
of moduli stabilization (see e.g.
\cite{Becker:2001pm,Dasgupta:1999ss,Gukov:1999ya,GKP,Silverstein:2001xn,Maloney:2002rr,KKLT}, as well as the
reviews \cite{Frey:2003tf,Silverstein:2004id,Grana:2005jc,PolchinskiCC,Douglas:2006es,Denef:2007pq,Boussoreview}
and references therein) has shown that in many classes of compactifications, the four-dimensional effective
action contains a rich set of potential energy terms that plausibly contribute the requisite forces
\cite{GKP,Silverstein:2001xn,Maloney:2002rr,KKLT}.  This realizes the possibility of a ``discretuum" of vacua
introduced in \cite{Bousso:2000xa} (see also \cite{FMRSW}).

Some earlier developments in string cosmology relied on the hope that whatever mechanism eventually stabilized
the moduli would not have side effects for models of inflation and dark energy.  We know now that this hope,
though not unreasonable, is violated in a wide class of models: the details of moduli stabilization do have
important effects in cosmological models.  Nevertheless, moduli stabilization can now be studied in
semi-explicit constructions, and this gives current models a certain degree of completeness and realism.

This confluence of observational and theoretical advances makes the present situation rather exciting. Even if
there are no further surprises, observational advances will certainly zero in on a tiny fraction of the present
parameter space; moreover, the techniques exist to make relatively robust models of the resulting phenomena in
string theory.

In this review we will not attempt to give a comprehensive assessment of the status of inflation, let alone of
more general cosmological model-building.  Indeed, the recent period has seen a number of useful reviews of the
subject, which complement ours in some ways \cite{reviews}; this leaves us the freedom to comment most
extensively on the aspects of string cosmology with which we are most familiar. It bears mentioning that,
although inflation is by far the most compelling explanation for the observed properties of the very early
universe, its status is the subject of some debate, and most authors find that some degree of fine-tuning
appears to be necessary. Thus, it is worth keeping in mind the possibility of an entirely different solution to
the Big Bang problems.

More generally, there are important questions in string cosmology that inflation renders essentially
inaccessible experimentally, such as the resolution of spacelike singularities, or phase transitions connecting
different limits of string theory. Microscopic physics is clearly relevant to these questions, which are
motivated by intellectual interest regardless of near-term testability.\footnote{Perhaps it is worth emphasizing
in this regard that at the time that inflation was proposed, there was little or no confidence in the
possibility of measuring its consequences; one can hope that similar surprises in observational accessibility
will arise in the future.} Moreover, the possibility remains open that their resolution could introduce
consistency conditions or other effects relevant also at later times, perhaps as initial conditions for
inflation.  We will therefore include a discussion of these issues in \S2, and then move on to a survey of some
attempts at realistic model-building within string theory. Before moving to these developments, however, let us
flesh out the basic reasons for the mutual relevance of string theory and cosmology.

\subsection{String theory and cosmology}

Inflation provides a solution to the flatness and horizon problems within the framework of quantum field theory
(QFT) and general relativity (GR).  Moreover,
in so doing it screens from observational view much of the physics of the ultraviolet (UV) completion of this
framework, whatever that might be.  Indeed, inflation explains the {\it absence} of observable relics, such as
monopoles from possible extensions of the Standard Model. One may wonder, therefore, whether string theory, as a
candidate UV completion of particle physics and gravity, should play much of a role in cosmology at all.
However, there are several reasons for greater optimism, to which we now turn.

\subsubsection{Utility of string
theory in theoretical cosmology}

There are several theoretical motivations to incorporate string theory into cosmological model building.  First,
the observed expansion of the universe, extrapolated backward in time using the equations of GR and QFT,
inevitably hits a regime where these descriptions break down \cite{Borde:2001nh}.  The corresponding spacelike
singularity requires physics beyond the Standard Model and GR to resolve.  For this reason, and also because of
the causal structure of spacetimes with inflation and dark energy, we lack a coherent framework for analyzing
cosmology from start to finish.
Second, there is the possibility that the degrees of freedom appropriate to describe
cosmological spacetime could be understood at a more fundamental level by enumerating microscopic degrees of
freedom that account for the Gibbons-Hawking entropy, in the same way that black holes have been illuminated by
the determination of the microscopic origin of the Bekenstein-Hawking entropy \cite{StromingerVafa} in certain
cases.  Finally, at a more practical level, many field-theoretic cosmological models involve ranges of energy
scales and scalar field strengths which reach large enough values that generic string or Planck-scale
corrections become significant.

\subsubsection{Cosmology as a
test of string theory}

Conversely, cosmology is a natural place to seek concrete tests of string models. First of all, purely in
comparison to collider experiments, cosmological observations hold greater promise for testing string theory.  A
natural energy scale for inflation is around the GUT scale, and an observation of primordial tensor fluctuations
in the CMB would establish beyond doubt that the CMB fluctuations were generated at such energies. No planned
terrestrial experiment can reach a fraction of this energy.  Hence, the decoupling of high-energy phenomena
strongly suggests that unless by some strange chance the string scale is extremely low, signals of string theory
will be seen, if they are seen at all, in the sky.

This must not be taken to mean that CMB experiment is either a direct or a guaranteed probe of string theory.
Almost any signal that can arise in a string model can arise in a suitable low-energy effective quantum field
theory.  Distinguishing these possibilities is highly nontrivial and could be impossible in many cases.  With
this in mind, there are still two avenues for a possible test.  In fortunate circumstances, some phenomenon does
not fully decouple at low energies; cosmic strings, which are topological defects, are an important example.
Another way to use cosmological observations to constrain string theory is to check for signals which are
natural or generic in string-derived effective Lagrangians, but highly unnatural from a conventional
field-theory viewpoint. Observation of such a signal would be suggestive, at the very least.  A specific form of
this strategy is to find signals that {\it cannnot} arise in effective field theories derived from certain
string theory constructions.  As an example, in many -- but {\it{not}} necessarily all -- string inflation
models, the primordial tensor signal is very small, as we review in \S\ref{subsubsec:PlanckRange}.  Hence, an
observation of primordial tensors would eliminate the great majority of presently-known string inflation models.
Eliminating classes of string constructions in this way does not amount to a test of string theory itself, but
would nevertheless play a valuable role by restricting the currently formidable range of possibilities of the
theory.

In this spirit, the fact that inflation models require control over the action at a level sensitive to
Planck-scale corrections (as in the supergravity eta problem reviewed in \S\ref{subsec:eta}) may provide
information about such microphysical effects, at least in specific frameworks for model building.  Of course,
with only a small number of operators constrained in this way, the problem of tying observations to string
theory is still enormously underdetermined.

\section{The Initial Singularity in String Theory}

A UV completion of GR is most directly relevant at very early times, and becomes crucial at the initial
singularity.  Our understanding of the resolution of spacelike singularities is very limited, but recent work
has shown how the new degrees of freedom introduced by string theory play a significant role, plausibly
resolving the singularity in special examples.  Without embarking on a comprehensive review of this large
subject (see \cite{Garytalk,Berkooz:2007nm,Craps:2006yb,Cornalba:2003kd}\ for some recent lectures and reviews),
let us here mention some concrete ways in which string-theoretic microphysics intervenes on the way back to the
would-be initial singularity.

\subsection{Approaching the singularity in the effective theory}

The low energy field content of string theory affects the approach to a spacelike singularity as explained in
e.g. \cite{Damour:2002et}.  The inclusion of $p$-form fields as well as curvature leads to chaotic dynamics in
this regime, generalizing the BKL analysis of the approach to the singularity in four-dimensional general
relativity \cite{Belinsky:1970ew}\ (see e.g. \cite{Garfinkle:2003bb}\ for recent numerical studies). The fields
bounce around in a potential consisting of a sum of exponential walls determined by the fluxes and curvature
terms, leading to a billiards problem; this is related to the fact that upon dimensional reduction of a gravity
theory, the potential energy is exponential in the canonically normalized scalar fields descending from the
metric and other modes. In a realistic setting, the number of times the field bounces off a wall is limited, but
conceptually the effect is quite interesting and suggests the possibility of an enhanced symmetry near the
singularity.

Another important feature of many cosmological models is a regime of eternal inflation at very early times. A
version of this process that is controlled within effective field theory arises in the presence of metastable
vacua, which are a feature of string-theoretic models fixing moduli to obtain (metastable) de Sitter space
\cite{Maloney:2002rr,KKLT}, described further in \S6 below.  Attempts to describe eternal inflation
holographically, or to determine a probability measure in this context, include
\cite{holography,Maldacena:2002vr,measures}.  Two of the main issues are as follows.  First, heuristically one
might expect the most rapidly-inflating region to give the dominant probability, because of its exponentially
larger volume.  However, this description is not automatically gauge invariant in general relativity, and in
addition there are limits to predictivity in any closed quantum mechanical system. Second, in the regime
described well by effective field theory and GR, the causal structure of an eternally inflating spacetime is
such that no single observer can access the full geometry. Operationally, then, it might make the most sense to
consider a formulation based on the causal patch of a single observer. Various proposals addressing these issues
have been put forward, and the subject is an active area of current research.  Once a mathematically
well-defined framework is obtained, one would like to know if it is uniquely specified, what predictions it
makes, and whether or not the results are sensitive to initial conditions.\footnote{For recent works exploring
dynamics on the landscape, see e.g. \cite{tunneling}.}

\subsection{Perturbative string theory and winding states}

Eventually, a treatment based on QFT and GR breaks down, either when the Hubble expansion rate increases to the
scale of some new physics, or when features in the spatial three-manifold shrink to the string scale (in a
background with weak string coupling $g_s\ll 1$) or the Planck scale (in a background with $g_s\sim 1$). In
particular, perturbative string degrees of freedom become important in large classes of examples. Consider the
spacetime \be ds^2=-dt^2+a^2(t)ds^2_{{\cal M}_3}+ds^2_\perp \ee where ${\cal M}$ is some three-manifold.
Almost all compact spatial manifolds ${\cal M}_3$ -- in fact\footnote{One of the major developments of recent
years being the proof of the Poincar\'e conjecture by mathematicians, using some techniques from physics.} all
but the three-sphere $S^3$ -- include topologically nontrivial one-cycles, which support new single-string
states corresponding to strings wound around these cycles. Indeed, there is a well-defined mathematical sense in
which most compact manifolds are hyperbolic \cite{Thurston}\ and hence have a fundamental group of exponential
growth \cite{Milnor}.\footnote{Inflation stretches ${\cal M}_3$ exponentially, but the interesting logical
possibility that its topology is accessible just within our horizon has been studied and bounded, e.g. in
\cite{BondHyperbolic}.} That is, the number of closed geodesics as a function of their length $l$ grows
exponentially with $l$:
\be \rho(l)\sim e^{l/l_0} \label{expgrowth}\ee
where $l_0$ is of order the curvature radius, and we neglect power-law prefactors. Thus, since winding string
masses scale like $l/\alpha'$, there is generically an exponentially growing (Hagedorn) density of winding
strings as a function of their mass. As space shrinks in the approach toward the singularity, these states,
among others, become important to the dynamics, and they enhance the effective dimension of the system, as
follows. In string theory in flat space, there is a an exponential density of states arising from oscillations
of the strings in the dimensions on which they propagate; for larger dimensionality, the density of states grows
more rapidly. More generally, one can extract the effective dimension (or ``effective central charge") from the
high-energy density of states in any perturbative string background \cite{Kutasov:1990sv}. At large but finite
radius, this Hagedorn density of winding strings on ${\cal M}_3$ leads to a calculable enhancement to the
effective central charge, precisely consistent with modular invariance, indicating a supercritical string
spectrum in the approach to the singularity \cite{Dduality,Aharony:2006ra}. This raises the question of what
happens at smaller radius, for example in the window where string $\alpha'$ effects become important but $g_s$
is still small.

Let us begin with the case where spacetime fermions have periodic boundary conditions around the $b_1$ homology
one-cycles in ${\cal M}_3$.  In a particular controlled limit one can show that the minimal possible completion
within perturbative string theory, preserving the symmetries of the system, is a $b_1$-dimensional torus
\cite{Dduality}. In other words, a rudimentary study of generic cosmological spacetimes in string theory reveals
a novel form of string duality (dubbed D-duality): new dimensions emerge from topology. This provides simple,
well-motivated examples in which the effective central charge decreases from a supercritical value to the
critical dimension -- the final stages of the transition proceeding simply by Hubble expansion, which reduces
the effective central charge by decreasing the contribution of the semiclassical winding strings. This
establishes, in a regime with a well-controlled expansion in both $\alpha'$ and the string coupling, that the
supercritical string is dynamically closely connected to the critical limits of string theory. Recent progress
in the works \cite{HellermanSC}\ links highly supercritical and critical corners of string theory, using very
useful new techniques involving tachyon condensation along a lightlike surface -- a special configuration which
affords exact control of $\alpha'$ effects and can be applied to connect many disparate limits of string theory.
We will discuss the generic higher-dimensional phases of string theory further in \S5.2.

In the above cases, the string coupling ultimately grows large, and a nonperturbative formulation is required.
However, perturbative string theory can sometimes be tractable and self-consistent on sufficiently simple
spaces. Many works, e.g.
 \cite{ChicagoLMS,Liu:2002yd,Cornalba:2003kd,Berkooz:2004re,Berkooz:2004yy}, have analyzed spacetime singularities
in time-dependent backgrounds involving circles with periodic boundary conditions for fermions, in order to
check if this analysis can be self-consistent, and if so to check whether string theory yields a smooth
crunch-bang transition. A basic challenge is to control the blueshifting of modes, whose energy density tends to
approach the Planck scale as the singularity is approached; these can change the character of the singularity,
for example rendering it spacelike rather than lightlike \cite{Horowitz:2002mw}.  Earlier work using worldsheet
conformal field theory techniques can be found, for example, in \cite{nappiwitten}\ (where the spatial slice is
an $S^3$).

A class of examples in which winding string physics intervenes to smooth out the singularity before the
blueshifting modes reach the Planck scale is obtained by considering one-cycles about which spacetime fermions
have {\it{antiperiodic}} boundary conditions -- and which shrink slowly, in string units, as the space reaches
the string scale in size: $H={\dot a\over a}\ll {1\over l_s}$.
%
%
A simple example of this in two dimensions is the Milne spacetime $ds^2=-dt^2+v^2t^2d\theta^2$ with $v\ll
1\Rightarrow {\dot a\over a}|_{vt=l_s}={v\over l_s}\ll {1\over l_s}$.

In this situation, the spectrum of strings can be reliably computed and includes a mode that becomes tachyonic
as the circle circumference diminishes to the string length. The mass-squared of the lightest strings wound
around a circle in ${\cal M}_3$ is
\be m^2 \sim \frac{L(t)^2}{l_s^4}-\frac{1}{l_s^2}+{\cal O}(H^2) \ee
where $L(t)$ is the proper length of the string and the second term comes from the negative Casimir energy on
the string worldsheet.  This state becomes a {\it winding tachyon} $T$ for $L(t)<l_s$, and the condensation of
this mode drives the system in a different direction from the singularity predicted by general relativity
\cite{McGreevy:2005ci}.  At weak string coupling, this effect arises parametrically far from the regime of
Planck-scale physics, since the circle is of size $l_s\sim l_{pl}/g_s\gg l_{pl}$ (where $l_{pl}$ is the Planck
length), and $H$ is parametrically small compared to the string scale.

This condensate produces a spacetime mass term in the sigma model for strings propagating on this spacetime
\cite{Strominger:2003fn}, proportional to a relevant operator in the spatial directions of the worldsheet sigma
model. This suggests that the tachyon effectively masses up all the degrees of freedom of the system.  This
intuition is borne out by calculations of the leading time-dependent effects in the system, made using the ``in
vacuum" in the tachyon phase. Namely, the particle production read off from the two-point function is exactly
the same as in a field-theoretic system with exponentially growing mass, and the one-loop vacuum energy gets
contributions only in the region of spacetime in which the tachyon has not condensed.  This scenario therefore
yields a perturbative string mechanism for an old idea in Euclidean quantum gravity -- that of starting time
from ``nothing".  The tachyon dynamics is quite subtle; one open question concerns the set of allowed states and
how to formulate the worldsheet theory appropriate to states other than the in vacuum.

\subsection{Nonperturbative formulations and the singularity}

In many cases, such as rapidly shrinking spaces and spaces with periodic boundary conditions for fermions,
nonperturbative effects may ultimately be required to resolve the singularity. It is possible to formulate
string theory nonperturbatively in various special spacetime backgrounds, using matrix theory
\cite{Banks:1996vh}\ or the AdS/CFT correspondence \cite{Aharony:1999ti}.

These formulations are not well-adapted to cosmology as it stands,\footnote{AdS/CFT requires a timelike
boundary, while Matrix Theory is based on graviton scattering in flat spacetime, and has not been generalized to
the case of compactification down to four-dimensional target spaces.} but can be used to gain insight into some
null and spacelike singularities. Investigations along these lines have suggested methods for calculating
effects of the singularity from the gauge theory side
\cite{Kraus:2002iv,Fidkowski:2003nf,Hertog:2004rz,HongAdSsing}\ and in some cases yield a set of free Yang-Mills
degrees of freedom associated with the singularity \cite{Craps:2005wd,Das:2006pw,Dduality}.

\section{Inflation in Field Theory and String Theory}

Although string theory is most directly relevant in the extreme conditions near the initial singularity,
inflation, if it occurred, is very effective at screening this microphysics from observation.  However, traces
of the inflationary epoch itself do remain accessible, and may be our best hope to probe physics at very high
energies.  Let us therefore turn to inflationary model building and its interface with string theory.

\subsection{Inflation in quantum field theory}

Inflation provides a solution to several cosmological problems within the realm of effective field theory. We
will consider Lagrangians of the fairly general form \cite{Garriga:1999vw}

\begin{equation} \label{general}
S=\int d^4x \sqrt{-g} \left[\frac{1}{2}M_{pl}^2{\cal R} + P(X^{IJ},\phi^J)\right]~,
\end{equation}
where $\phi_I$ are scalar fields including the inflaton field,
$X^{IJ}=-\frac{1}{2}g^{\mu\nu}\partial_{\mu}\phi^I
\partial_{\nu}\phi^J$, and $P$ is some function. The reduced Planck mass is $M_{pl}=(8\pi G)^{-\frac{1}{2}}$.

Two familiar special cases are the two-derivative action for a general curved field space, with
\be \label{SREFT} P_{SR}=-G_{IJ}\del\phi^I\del\phi^J-V(\phi)\ee
for some metric $G_{IJ}(\phi^K)$, and the DBI action
\be \label{DBIEFT} P_{DBI}=-f(\phi)^{-1}\sqrt{1+{f(\phi)(\del\phi)^2}}-V(\phi) \ee
where $(\del\phi)^2 \equiv g^{\mu\nu}\del_{\mu}\phi\del_{\nu}\phi$, and we have suppressed the $I,J$ indices.
Here $f(\phi)$ is a function that, although computable in concrete string theory constructions, may be taken to
be general for the present purpose.  Each of these actions is valid in the limited regime in which all neglected
operators, evaluated on the solution of interest, are subdominant to the terms listed.  The question of what
terms dominate can be quite subtle and sensitive to the details of the UV completion, and this is the source of
some of the difficulty, and utility, of embedding these scenarios into string theory. (An effective field theory
treatment appropriate to expansion about inflating backgrounds was developed very recently in
\cite{InflationEFT}.)

We consider approximately homogeneous and isotropic solutions given by the Friedmann-Robertson-Walker metric
\begin{equation}
ds^2=-dt^2+a^2(t)dx_3^2 ~.
\end{equation}
The Hubble expansion is characterized by $H=\frac{\dot{a}}{a}$ and is determined by the sources of stress-energy
via the Friedmann equations. The stress-energy tensor arising from rolling scalar fields governed by
(\ref{general}) has pressure $P(X,\phi)$ and energy density
\begin{equation}
\rho=2X P_{,X} -P ~, \label{EPdef}
\end{equation}
where $P_{,X}$ denotes the derivative with respect to $X$.

Much recent work has focused on embedding slow roll inflation, an extensively studied mechanism for inflation in
quantum field theory based on the two-derivative action, into string theory. At the same time, string theory --
in particular the dynamics of scalar fields in AdS/CFT dual pairs -- has led to the development of a new
field-theoretic mechanism for inflation, obtained from the effective action (\ref{DBIEFT}).  The resulting DBI
inflation scenario would benefit from an explicit UV completion, and work continues in this direction. We will
discuss progress toward constructions along both these lines in \S\ref{sec:InflationModels}.

In any known field-theoretic mechanism for inflation, some fine-tuning of Lagrangian parameters is required.
That is, the effective action (\ref{general})
must be chosen carefully to arrange for a long period of inflation producing a suitable spectrum of density
perturbations; natural, generic values for these data do not imply a solution of the Big Bang problems. One of
the more conservative goals of the subject is to determine explicitly how to implement this tuning in a
UV-complete framework, or, conceivably, to discover correlations in couplings that might descend from a full
theory even if they appear tuned in the IR theory.\footnote{See \cite{BDKM} for a recent attempt along these
lines in the specific context of D-brane inflation.}

Specifically, one can define (generalized) slow variation parameters whose smallness ensures an extended period
of inflation (discussed e.g. in \cite{Chen:2006nt}) as follows:
\begin{eqnarray} \label{small}
\epsilon&=& -\frac{\dot{H}}{H^2}=\frac{XP_{,X}}{\Mpl^2 H^2}~, \nonumber \\
\tilde\eta &=& \frac{\dot{\epsilon}}{\epsilon H}~, \nonumber \\
s &=& \frac{\dot{c_s}}{c_sH} ~,
\end{eqnarray}
where the sound speed $c_s$ is given by
\begin{eqnarray}
c_s^2 = \frac{dP}{d\rho}= \frac{P_{,X}}{P_{,X}+2XP_{,XX}} \, .
\end{eqnarray}

In the case of slow roll inflation (\ref{SREFT}), one ensures slow variation by choosing an appropriately tuned
flat potential $V(\phi)$ which one must control (along with the rest of the action) over a sufficient range in
field space.  In the case of DBI inflation (\ref{DBIEFT}), the slow motion of the canonically normalized scalar
field is ensured by the ``speed limit" enforced by the square root action, ensuring that $\dot\phi^2<1/f(\phi)$
regardless of $V(\phi)$; one must choose an appropriately tuned $f(\phi)$ and control it (along with the rest of
the action) over a sufficient range in field space.

The Gaussian approximation to the scalar power spectrum derived from the two-point function is given by
\cite{Garriga:1999vw}
\begin{equation} \label{power1}
{\cal P}^{\zeta}_k
=\frac{1}{8 \pi^2 M_{pl}^2}\frac{H^2}{c_s\epsilon} ~,
\end{equation}
where the expression is evaluated at the time of horizon exit at $c_s k=aH$.  Observations give ${\cal
P}^{\zeta}_k\approx 10^{-9}$. The spectral index is
\begin{equation} \label{index1}
n_s-1=\frac{d\ln {\cal
P}^{\zeta}_k}{d \ln k}= -2\epsilon-\tilde{\eta}-s ~.
\end{equation}
The power in tensor perturbations at the Gaussian level is \beq \label{tensorpower} {\cal P}^{h}_k =
\frac{2}{\pi^2}\frac{H^2}{\Mpl^2} ~. \eeq The tilt and tensor-to-scalar ratio $r={\cal P}_h/{\cal P}^\zeta$ are
bounded such that $n_s\sim 0.95$ at $r\sim 0$ while larger $r$ correlates with the possibility of larger $n_s$
\cite{Observations}.

The possibility of detecting gravitational waves from inflation \cite{Starobinsky} is in fact an example of a
question which is very sensitive to UV physics. Lyth has shown that a detectable gravitational wave signal
requires a super-Planckian field range for the canonically normalized inflaton field \cite{Lyth}.  This extends
to more general theories of the form (\ref{general}) \cite{ShamitTalk}.  Arranging for the slow variation
parameters (\ref{small}) to be small over such a range is extremely delicate: corrections ${\cal
O}_\Delta/\Mpl^{\Delta-4}$ to the effective action involving higher dimension operators ${\cal O}_\Delta$ of
dimension $\Delta>4$ generically become important if the inflaton ranges over a Planck-scale (or higher) range
in field space. Moreover, in most controlled backgrounds of string theory there are degrees of freedom at a
lower scale than $\Mpl$, leading to larger effects that interfere with inflation even over sub-Planckian field
ranges.

The Lagrangian (\ref{general}) contains higher-point interactions of the perturbations of $\phi$ and the
graviton. Therefore there is a non-Gaussian correction to the power spectrum.  In single-field slow roll
inflation, the leading such effect includes couplings to gravitons and is unmeasurably small (see
\cite{Maldacena:2002vr}).  The inclusion of higher-dimension operators suppressed by a high mass scale as in
\cite{CreminelliHD}\ improves the prospects somewhat. Building on this, more generically one obtains
higher-dimension operators suppressed not by a hard mass parameter, but instead by the inflaton vev itself,
$\phi$ \cite{Silverstein:2003hf}.  In this circumstance the non-Gaussian correction can be substantial
\cite{Alishahiha:2004eh}.   As an example, in DBI inflation (\ref{DBIEFT}), expanding in perturbations about the
background solution brings down powers of $\gamma=1/\sqrt{1-f\dot\phi^2}$, which is large in the solution when
$\dot\phi$ approaches its speed limit. This correction is large enough to ensure that the DBI inflation
mechanism is falsifiable:  its entire parameter range will be covered by upcoming experiments.\footnote{This
statement is based on the original predictions of the sensitivity of the Planck satellite, for example, which
may be revised downward.} The non-Gaussian correction to the power spectrum has been computed in general
single-field inflation in \cite{Chen:2006nt}\ following \cite{Maldacena:2002vr,CreminelliHD,Alishahiha:2004eh};
an effective field theory treatment appears in \cite{InflationEFT,InflationEFTconsistency}. To first order in
the slow variation parameters, the three-point function for the gauge-invariant scalar perturbation $\zeta$ is
of the form
\begin{eqnarray}
\langle \zeta(\textbf{k}_1)\zeta(\textbf{k}_2)\zeta(\textbf{k}_3)\rangle &=&
(2\pi)^7\delta^3(\textbf{k}_1+\textbf{k}_2+\textbf{k}_3) (\tilde {\cal
P}^{\zeta}_k)^2 \frac{1}{\prod_ik_i^3} \cr
&\times& (\CA_\lambda +{\CA}_c + \CA_o +\CA_\epsilon +\CA_\eta +\CA_s) ~. \label{3pointFinal}
\end{eqnarray}
The shape decomposes into the six functions ${\cal A}_I, I=1,\dots,6$ given in Appendix B1 of
\cite{Chen:2006nt}.  Large non-Gaussianity is correlated with small sound speed; in DBI inflation $c_s\sim
1/\gamma$.

Non-Gaussian corrections have been bounded by the WMAP data \cite{Observations}\cite{HarvardNG}.  The
three-point function (\ref{3pointFinal}) depends on two independent momentum vectors, or equivalently on a
triangle in momentum space. The amplitude of the effect as a function of momentum triangles depends on the
model; for DBI inflation the WMAP data currently bounds  $\gamma \lesssim 20$, for example. There are currently
investigations into other observational methods to bound (or detect) primordial non-Gaussianity.

Some authors have explored the question of whether there is a nontrivial choice of vacuum for the scalar
perturbations (deviating from the Bunch-Davies vacuum assumed above), and if so how this would affect the power
spectrum (see for example \cite{Easther:2002xe,Kaloper:2002uj,Kaloper:2002cs}). The conservative view seems to
be that an initial excitation above the Bunch-Davies vacuum would thermalize during inflation, reducing the
system back to Bunch-Davies; however, even given this there is a chance that initial transients appear within
the CMB window of observability. A nonstandard choice of vacuum endows (\ref{power1}) with a momentum-dependent
prefactor which can be bounded by observations.

In addition to probes of and constraints on inflation from the power spectrum, there are observational
constraints and opportunities from relics such as cosmic strings produced in the exit from inflation. We will
not cover this in detail since it has been reviewed very recently in
\cite{Polchinskicosmic,Tye:2005wv,Polchinski:2007qc}.  In the context of string theory, the gravitational
redshift (warping) used in the scenarios of \cite{KKLMMT} (\S4.2) to obtain inflation with suitably small ${\cal
P}_k^\zeta$ has the added effect of redshifting down the string tension in D-brane inflation
\cite{Sarangi:2002yt}\ to a viable scale \cite{Copeland:2004iv}.  This excess of outputs over inputs is one of
the encouraging signs in the subject as developed thus far.  The cosmic superstrings arising in this context
have a rather rich phenomenology \cite{MoreCosmicStrings}. An important step was made in realizing that cusps
and kinks on cosmic strings lead to a distinctive gravity wave signature \cite{DamourVilenkin}.


\subsection{Challenges for inflation in string theory}

The primary task in string inflation is to specify a string compactification whose low-energy effective
Lagrangian is capable of producing inflation that is consistent with current observations.  This turns out to be
a surprisingly difficult problem, and there is no definitive proof of a {\it{fully}} successful model at the
time of this writing. Once this has been achieved, one can search for special features of the string-derived
Lagrangian that might provide characteristic signatures of the model. In this section we will explain the
obstacles to success, and in the next section we will describe a range of relatively successful models. Of
course, as discussed already, string theory also contributes to the development of inflation at the
field-theoretic level by inspiring new mechanisms and models, some of which are testable independently of their
UV completion.

\subsubsection{From compactification data to the inflaton Lagrangian}

In principle, one would like to construct a string inflation model by specifying the data of a consistent string
compactification, including the total dimensionality; the geometry and topology (or generalization
thereof)\footnote{Although here we will focus mostly on geometric backgrounds, various authors have emphasized
the possible genericity of non-geometrical ones \cite{Wecht}, and indeed early landscape models made use of
asymmetric orbifolding, and hence were of this more general type \cite{Silverstein:2001xn,Maloney:2002rr}.}; the
locations of any D-branes, orientifold planes, and other localized sources; and the amount of flux turned on
through each cycle. Such a configuration would uniquely specify a four-dimensional classical Lagrangian, and our
knowledge of this theory would be limited only by the accuracy of the dimensional reduction -- for example, by
$\alpha^{\prime}$ and $g_s$ corrections, or by backreaction effects from the localized sources.\footnote{Of
course, the quantum-corrected value of the cosmological constant cannot be computed in this setup, and in any
framework for modeling the real-world dark energy, a genericity argument is used to argue that the quantum
corrections can be cancelled by appropriate tuning of discrete quantum numbers.}

At present, however, most models are constructed less directly, through a certain amount of informed guesswork.
For example, moduli stabilization is most often accomplished by invoking a general method, such as the proposal
\cite{GKP,KKLT}, instead of by a precise specification of fluxes and perturbative and nonperturbative effects;
ingredients responsible for tadpole cancellation, such as orientifold planes, are assumed to be so far away from
the region where the inflation effect is localized that they have no effect on the potential; the metric on a
noncompact cone is sometimes used as an approximation to that of a finite conical region in a compact space; and
very often $g_s$ and $\alpha^{\prime}$ corrections are omitted without sufficient concern for their effects on
the putative inflationary solution.

In many cases, these approximate treatments are the best that can be accomplished with present knowledge.
Approximations of this sort suffice only as long as they do not lead to important inaccuracies, which in the
worst case can cause a model to appear successful at leading order but to fail totally when corrections are
included. Although at present no entirely systematic method exists for estimating the errors in all these
approximations, we will present, in \S\ref{subsec:eta}, a useful scheme for organizing corrections to the
leading-order dimensional reduction. This helps to identify cases in which a model is in jeopardy and
computation of higher corrections is necessary. To justify the scheme, however, we first explain two general and
important obstacles in string inflation model-building: the moduli problem and the eta problem.

\subsubsection{The moduli problem and the eta problem}
\label{subsec:eta}

The first difficulty is that string compactifications invariably involve more than one scalar field, and so are
much more complicated than the single-field Lagrangians of the preceding section.  The four-dimensional
potential depends, in general, on all the moduli of the compactification, which parametrize the geometry of the
internal space.  In Calabi-Yau compactifications, one has K\"ahler moduli, complex structure moduli, and the
dilaton, numbering in the hundreds in typical cases.  Computing the full potential as a function of all these
fields is a formidable task, and even if one could succeed, it would then be necessary to search through the
high-dimensional field space for a path along which the resulting potential\footnote{For simplicity of
exposition, we focus here on realizing slow roll inflation,
 but the issues are analogous for more general models, in which the potential need not be flat but other operators
 in the action are constrained by the need for small generalized slow roll parameters (\ref{small}).} is
 sufficiently flat for inflation.

An essential point is that it does {\it{not}} suffice to hold fixed, by hand, all fields but one, and then find
a path along which the potential for that single field, $\phi_1$, is flat.  One reason is that the full
potential will typically have a steep downhill direction coinciding with one or more of these other fields,
$\phi_2, \ldots \phi_N$.  If the steepest such direction is more steep than the desired, and nearly flat,
inflaton direction, the full system will evolve by rolling downhill in this steepest direction rather than along
the putative inflaton direction $\phi_1$. Thus, one must actually arrange that the potential has positive
curvature in the directions of $\phi_2, \ldots \phi_N$.  This, too, is not sufficient: a field $\chi$ with $ 0 <
m_{\chi}^2 < \frac{3}{2}H^2 $ will undergo quantum fluctuations during inflation.  These fluctuations carry the
field away from its minimum and hence lead to storage of energy in $\chi$.  If $\chi$ is quite light and couples
only gravitationally (as moduli do), the $\chi$ particles will not have decayed by the present day, and will
overclose the universe.  If $\chi$ is somewhat heavier, up to $\sim 30 ~{\rm{TeV}}$, it will have decayed during
or after Big Bang Nucleosynthesis, spoiling the delicate predictions of the light element abundances.  To avoid
this `cosmological moduli problem', one must\footnote{Axions with sufficiently small periodicity can be viable
below this bound, as their potential energy and initial displacement from their minimum are smaller than for
generic moduli; see \cite{Svrcek:2006yi}\ for a recent discussion of the constraints on axions and a survey of
their parameters in a sample of string backgrounds.} therefore arrange that $ m_{\chi}^2 > 30 ~{\rm{TeV}}$, and
preferably that $m_{\chi}^2 \gg H^2$.  Giving moduli large positive masses is known as moduli stabilization.


Explicit techniques that stabilize most or all of the compactification moduli became available only in the last
decade.  An approach pursued in the absence of moduli stabilization was to assume that whatever mechanism
stabilizes the moduli does not affect the inflaton potential.  This turns out {\it{not}} to be a valid
assumption.  One qualitative reason to expect a problem is that moduli stabilization aims to arrange
$m_{\chi_i}^2 \gg H^2$ for all the runaway moduli $\chi_i$,\footnote{It is possible for moduli other than the
inflaton to have lighter masses, as long as they sit at a local minimum during inflation and as long as they sit
at or roll toward a stable minimum at the end of inflation.} whereas inflation requires $m_{\phi}^2 \ll H^2$.
Since the inflaton is usually nothing more than a carefully-chosen modulus, the origin of this sharp disparity
is hard to justify: whatever mechanism lifts the flat directions of all the other moduli potentials will also
lift the inflaton flat direction.  This expectation has now been established in a wide range of examples, and is
in fact a specific incarnation of a more general difficulty known as the eta problem.

We first recall that from the definition of the slow roll parameter $\eta$, \beq \eta \equiv
\Mpl^2\frac{V^{\prime\prime}}{V} \, ,\eeq with primes denoting derivatives with respect to an inflaton $\varphi$
with a canonical kinetic term, we have $m_\varphi^2 = \frac{V}{\Mpl^2}\eta = 3 H^2 \eta$.  Hence, the inflaton
mass being of order $H$ is equivalent to having $\eta \sim 1$.  Moreover, as $\eta$ governs the duration of slow
roll inflation, $\eta \sim 1$ leads to an unacceptably small amount of inflation.

The eta problem is the observation that, in string theory and effective field theory realizations of inflation
(including supersymmetric ones), Planck-suppressed corrections to an otherwise flat inflaton potential
generically give rise to inflaton mass terms of order $H$, leading to $\eta \sim 1$, and hence ruining slow roll
\cite{CopelandEta}. Specifically, Planck-suppressed contributions to the potential of the form \beq \Delta V =
\frac{{\cal O}_4}{\Mpl^2}\phi^2 \, , \eeq for some operator ${\cal O}_4$ of dimension four, lead to mass terms
\beq \Delta m_{\phi}^2 \propto \frac{\langle {\cal O}_4 \rangle}{\Mpl^2} \, . \eeq Then, if $\langle {\cal O}_4
\rangle \sim V$, these contributions lead to $\Delta \eta \sim 1$.

In the context of models with a low-energy supergravity description, the most transparent contribution of this
sort is present whenever there is an F-term energy of order the inflationary energy.  This condition is
obviously met when inflation is driven by an F-term, but is also typically obeyed when moduli stabilization is
accomplished via an F-term.  The intuitive reason for this is that a small moduli-stabilizing energy cannot
prevent a large inflationary energy from driving a runaway decompactification, so what is often obliged to
consider moduli-stabilizing energies of order the inflationary energy.\footnote{This relation could be violated
by an unusually steep moduli-stabilizing potential.}  In any such case, in terms of the K\"ahler potential $K$
and superpotential $W$, we have \beq \label{equ:vf} V \approx V_F = e^{K/\Mpl^2}\left(K^{A\bar{B}}D_A W D_{\bar B}\overline{W}
-\frac{3}{\Mpl^2} |W|^2 \right) \eeq Next, we expand $K$ around some fiducial origin $\phi=0$ as
$K=K(0)+K_{,\phi\bar{\phi}}(0)\phi\bar{\phi}+\ldots$ and find, for small $\phi$, \beq \label{equ:vfe} {\cal{L}} \approx
-K_{,\phi\bar{\phi}}\partial\phi\partial\bar{\phi}-e^{K(0)/\Mpl^2}\left(1+\frac{1}{\Mpl^2}K_{,\phi\bar{\phi}}\phi\bar\phi\right)\left(K^{A\bar{B}}D_A
W D_{\bar B}\overline{W} -\frac{3}{\Mpl^2} |W|^2 \right) + \ldots \, .\eeq Noting that the
canonically-normalized inflaton $\varphi$ obeys $\partial\varphi\partial\bar{\varphi}\approx
K_{,\phi\bar{\phi}}(0)\partial\phi\partial\bar{\phi}$, we see that the contribution to the mass term of
$\varphi$ is $\Delta m_{\varphi}^2 \approx V_F(0)/\Mpl^2 = 3 H^2$, so that $\Delta \eta = 1$.  In (\ref{equ:vfe}) we have indicated explicitly only the terms arising from the expansion of the exponential; the ellipsis stands in part for terms arising from expanding the factor in brackets in (\ref{equ:vf}).  These latter terms give contributions to $\eta$ of comparable size to those we display, but are more model-dependent.\footnote{For example, in the scenario of \cite{KKLMMT}, expanding $K$ everywhere it appears in (\ref{equ:vf}), not just in the exponential, one finds $\Delta\eta=2/3$.}

More generally, the eta problem will arise whenever a term in the potential of order $V$ is corrected
multiplicatively by a generic function of $\phi/\Mpl$.  For example, the nonperturbative superpotential used for
K\"ahler moduli stabilization in the KKLT scenario receives a multiplicative correction depending on the
positions of any D3-branes \cite{Ganor,BHK,GM,BDKMMM}.  In D3-brane inflation, this gives rise to a contribution
$\Delta\eta \sim 1$ \cite{KKLMMT,BHK,BDKMS,BDKM}, because, as mentioned above, the moduli-stabilizing energy is
generically of the same order as the inflationary energy.

An instructive way to organize corrections to the inflaton Lagrangian is in terms of the scale of the masses (or
other derivatives of the potential) they generate.  First, we denote by ${\cal L}_0$ the leading-order
Lagrangian arising from dimensional reduction.  Next, ${\cal L}_1$ is defined to include all corrections to
${\cal L}_0$ that introduce inflaton mass terms that are parametrically $\Delta m^2 \sim H^2$.  Finally, ${\cal
L}_2$ contains all corrections that introduce inflaton mass terms that are parametrically smaller than $\Delta
m^2 \sim H^2$, e.g. $\Delta m^2 \sim g_s H^2$ and the like.  In this language, merely to recognize the presence
of the eta problem, one must have some information about terms in ${\cal L}_1$; to be certain that the problem
is solved and the inflaton potential is very flat, one needs comprehensive information about ${\cal L}_2$.

In some cases, there is a concrete framework for computing these Planck-suppressed contributions to the
potential.  In ${\cal{N}}=1$ supergravity, these corrections more often appear in the K\"ahler potential than in
the superpotential, because the latter is protected by holomorphy.  String loop and $\alpha^{\prime}$
corrections to the K\"ahler potential then lead to the relevant terms in ${\cal L}_1$ and ${\cal L}_2$.
Unfortunately, computing these terms, especially those in ${\cal L}_2$, is extremely difficult and has only been
undertaken by a few authors \cite{BHKKahler}.

In the absence of such information, appeals to fine-tuning are sometimes made: one argues that in some
restricted subset of possible models, the net correction might be accidentally small.  The typical argument is
that the terms in ${\cal L}_1$ are not small, but happen to cancel each other to high precision for fine-tuned
values of the microphysical parameters.  This is an expectation, and has not been rigorously justified in most
cases.  In fact, a reasonably comprehensive examination \cite{BDKMS,BDKM} of the terms in ${\cal L}_1$ for
models of warped D-brane inflation has revealed that fine-tuning itself is possible only in a very restricted
subset of compactifications; in particular, there exist compactifications in which all the terms in ${\cal L}_1$
have the same sign.

Although the eta problem is well-known, one often encounters spurious claims that a given model of string
inflation is free of fine-tuning, when in fact it is simply the case that the potential has not yet been
computed with enough accuracy to ascertain whether the problem is present or not!  (Let us emphasize that we are
not suggesting that all such claims are incorrect in this way, only that many are.)

Another proposed solution to the eta problem is to include protective symmetries.  Although well-motivated, this
approach has been surprisingly difficult to achieve in explicit models, because the desired symmetries do not
always survive quantum corrections.

At present, the eta problem is one of the most serious constraints limiting our ability to construct explicit
models of string inflation.

\subsubsection{Constraints on field ranges}
\label{subsubsec:PlanckRange}

In a few models of inflation, the inflaton traverses a distance in field space that is large compared to the
Planck mass.  This category includes some of the scenarios which are simplest from the point of view of their
field content, such as Linde's chaotic inflation with a quadratic potential.  It also includes all models that
produce a detectably-large primordial tensor signal: as explained by Lyth, inflaton motion over a Planckian
distance is a necessary condition for such a signal.

To realize any such `large-field' \cite{Kinney} model in string theory, one would need to find a trajectory in
field space that is large in Planck units, and along which the effective action (\ref{general}) is suitable for
inflation. This has proved to be very difficult.  One way to anticipate this problem in the case of models with
a low-energy supergravity description is to write the possible corrections to the K\"ahler potential, \beq
{\cal{K}} = {\cal{K}}_{classical}(\Phi,\Phi^{\dagger}) + \Mpl^2 \sum_{i}
{c_i}\left({\Phi\Phi^{\dagger}\over{\Mpl^2}}\right)^{1+i} \eeq where the dimensionless coefficients $c_i$ may be
true constants or may depend on other fields in the system.  Unless the $c_i$ are all very small, this series is
badly divergent for $\Phi \gg \Mpl$, and so over Planckian distances, the metric on moduli space is
poorly-described by the classical metric on moduli space derived from ${\cal{K}}_{classical}$. More generally,
similar comments apply to the quantum-corrected effective action in cases without a low-energy SUSY description
in terms of a K\"ahler potential.

An exception arises near weak-coupling limits of string compactifications, which go off to infinite distance in
moduli space.  Here, the weak coupling allows one to calculate the effective action to a good approximation. At
least in the simplest examples yet studied, the result is an exponential potential in the canonically normalized
scalar fields, with a coefficient in the exponent which is of order one,
too steep for slow roll inflation. However, we are not aware of a sharp no-go theorem for inflation at this
level, in the generally rather complicated context of arbitrary compactification manifolds with many scalar
fields; it would be interesting to understand if there is a principled obstruction to inflation in tree-level
string theory.\footnote{For some explorations into the dynamics of multiple scalar fields descending from curved
compactifications, see for example \cite{Neupane:2005nb}.} Conversely, away from weak coupling it is difficult
to assert with confidence that one knows that the size of a given moduli space is large in Planck units.


In certain specific contexts in string theory, we can compute the field ranges more explicitly, in regimes
appropriate to candidate inflation models. An example of this is the field range for a D3-brane in a warped
throat region, a topic we will review in the next section. As shown in \cite{FieldRange}, for any sort of warped
throat arising from a cone over an Einstein manifold, the field range in Planck units is small. Similar bounds
are easily derived for D3-branes moving in (most) toroidal compactifications.  This result implies that D3-brane
inflation in Calabi-Yau throats, or in most tori, cannot give rise to an observably-large primordial tensor
signal.

Of course, nothing a priori requires D3-branes, and as explained very recently in \cite{BLS}, one interesting
way to evade this constraint is to develop an inflationary scenario involving the motion of a D($p+3$)-brane
wrapped on a $p$-cycle, for $p>0$. In such a case the relation between the geometric distance and the physical
canonical field is different than in the D3-brane case, and larger canonical field ranges are possible
\cite{BLS}. It is worth noting that in special compactifications, D3-branes {\it are} advantageous in the case
of slow roll inflation (described below in \S4.2.1) since their potential is flat at leading order by a
generalization of no-scale structure.  In the case of DBI inflation (described below in \S4.2.2), a steep
potential is useful, so the higher-dimensional branes may be practical in this case.

In closed string models, the field ranges correspond to distances in the space of geometric moduli, not
distances in the compactification itself.  An example of the situation discussed above of an infinite direction
in the moduli space is the decompactification direction.  In the case of low-energy supergravity, one
parameterizes this in terms of the \k\ potential and superpotential.  The \k\ potential depends on the total
volume ${\cal V}$ as \be {\cal K} = - 2 \Mpl^2 \log\left({\cal V}\right) \, , \ee so that $ R \equiv \Mpl
\sqrt{2} \log\left({\cal V}\right)$ has a canonical kinetic term.  The range of $R$ between any fixed ${\cal V}
\sim {\cal V}_{0}$ and the limiting point ${\cal V} \to \infty$ is arbitrarily large.  This would seem to be a
promising setting for large-field inflation, but it remains difficult to find a suitable inflaton potential
along this direction. The problem is related to the logarithmic form of the \k\ potential: energy sources that
have power-law dependence on ${\cal V}$ depend exponentially on the canonical field $R$ -- with a Planck-scale
coefficient in the exponent -- causing the potential to vary rapidly along the $R$ direction.

An interesting approach to constructing a microphysically-sensible super-Planckian vev was presented in
\cite{Nflation}.  By combining the displacements of $N \gg 1$ string axions into an effective displacement of a
collective field, these authors sidestepped the above restriction on individual field ranges.  However,
important questions remain about the radiative stability of this scenario, as we review in \S4.4.2.

Most successful closed string models involve potentials that are flat over very small ranges of the canonical
inflaton.  Such `small-field' models typically require fine-tuning and can be highly sensitive to initial
conditions, but they have the advantage that physics associated with Planckian displacements plays no role.

Notice that models with small field ranges, and hence \cite{Lyth} lower tensor signals, also have lower Hubble
scales during inflation, because the tensor signal (\ref{tensorpower}) is proportional to $H^2$.  In order for
inflation to set in, a smooth patch of linear size $H^{-1}$ is required.  Thus, small-field models require
smoother initial patches than large-field models require; on the other hand, large-field models require
{\it{functional}} fine-tuning in the effective action in order to allow for inflation over a super-Planckian
field range.  It would be interesting to understand whether the functional fine-tuning in large-field models is
counterbalanced by the necessity of smoother initial patches in small-field models.  This question is, in turn,
sensitive to earlier physics, including questions about probability measures from eternal inflation.

Another interesting, and rather strong, and constraint on the tensor signal in certain classes of string compactifications is \cite{KL}.

\section{Models of Inflation in String Theory}
\label{sec:InflationModels}

As we have discussed, proposing a model of inflation in string theory typically amounts to describing the data
of a compactification whose low-energy effective theory can contain a suitable inflaton field.  One often
requires as well that there is a plausible mechanism for reheating the universe.  The most potentially exciting
models are those with novel and distinctive signatures, but this is not a requirement by any means: it remains
worthwhile to construct definitively successful models of string inflation, even if these happen to lack
striking observational signatures, for precisely the same reason that is worthwhile to find string constructions
of the Standard Model of particle physics.

Because proposing a model involves identifying a scalar field as an inflaton candidate, we can classify string
inflation models according to the origin of the inflaton field.  Open-string models, usually called D-brane
inflation or brane inflation models, are those in which the inflaton is a scalar field arising from open strings
ending on a D-brane; some of the original papers on this approach are
\cite{Dvali:1998pa,Alexander:2001ks,Dvali:2001fw,SomeAspects,Burgess:2001fx,Burgess:2001vr,Jones}.  Usually this
scalar parameterizes transverse motion of the D-brane, and hence governs the location of the D-brane in the
compactification.  In M-theory, there is a closely-related alternative in which the inflaton corresponds to the
position of an M5-brane.

Closed string models are those in which the inflaton is a closed string mode.  The moduli are the most promising
closed string modes, as there is a well-defined choice of background for which a subset of them enjoy
classically flat potentials.  (Of course, this does involve a special choice of starting point; the {\it
generic} classical potential for moduli of the metric and $p$-form fields is too steep for inflation.) For this
reason, closed string inflation is sometimes called moduli inflation or modular inflation.  Classic works in
this framework include \cite{Oldmoduli}.

Because of limitations of space, we will not attempt to present a large fraction of current models here.
Instead, we will describe a few characteristic models in each category.  We do not mean to imply that the
scenarios selected are the most promising or most interesting among extant models; however, in our judgement
they are {\it{representative}} of the leading edge of contact between inflation and concrete string theory data.

It should be emphasized that any particular concrete model -- in field theory or in string theory -- is based on
specific, perhaps improbable, choices of microscopic ingredients and field content which are not forced upon us
directly by the data.
Nevertheless, it seems to us that investigations of this sort offer a surprising degree of insight into
important classes of behavior in string theory.  Moreover, models of the present generation have led to a range
of novel and genuinely predictive possibilities for inflation within quantum field theory.


\subsection{D-brane inflation}

D-brane inflation has been the subject of much recent work, and the result is a wide array of scenarios.  At
present D-brane models are best-studied in type IIB string theory, due in part to the comparatively better
understanding of moduli stabilization in the type IIB theory.\footnote{For example, the solutions can remain
conformally Calabi-Yau even in the presence of flux, which is an important technical advantage.}  One natural
candidate for the inflaton field is the position of a spacetime-filling D3-brane.  D3-branes feel no potential,
at leading order in $\alpha^{\prime}$ and $g_s$, in the no-scale flux compactifications of \cite{GKP}.  Thus,
the moduli space of a D3-brane in such a compactification is the entire internal space, and one can hope to find
a weak additional effect that will slightly lift this moduli space.  Examples of such effects include magnetic
flux on a D7-brane wrapping a four-cycle \cite{BKQ} (see also \cite{VZ}), an antibrane at the tip of a
Klebanov-Strassler throat \cite{KPV}, and a baryonic branch deformation of the Klebanov-Strassler geometry
\cite{DKS}, as well as $F$-term energy \cite{Fuplift}.

One interesting and well-studied proposal is D3-D7 inflation, in which a D3-brane moves towards one or more
D7-branes that wrap the $K3$ of a $K3 \times T^2/\mathbb{Z}_2$ compactification (or a more general space)
\cite{ThreeSeven}. This is a D-term inflation scenario, but F-term moduli stabilization (for example, by the
KKLT method), as well as mixing between the D3-brane moduli and K\"ahler moduli in the \k\ potential
\cite{DeWolfeGiddings}, lead to inflaton mass terms, as in \cite{KKLMMT}.  It was proposed
\cite{ThreeSevenShift} that geometric symmetries of the (toroidal part of the) compactification could protect
the inflaton potential from these problematic terms. However, one-loop threshold corrections to the
nonperturbative superpotential \cite{BHK} violate this shift symmetry and lead to a reappearance of the eta
problem in this context \cite{ThreeSevenMass}.  It would be interesting to understand whether there exists a
configuration of the D-branes in this compactification in which the inflaton mass term can be made to cancel.
Such a construction would still amount to a fine-tuning, not a natural solution of the eta problem, but would be
valuable nonetheless, given the high degree of computability of this model, and the access to explicit
information about moduli stabilization \cite{AspinwallKallosh}.

A more general possibility for the inflaton is a D($p$+3)-brane wrapped on a $p$-cycle.  In such cases, to study
the moduli space one needs information about families of $p$-cycles, which is easily obtained in toroidal and
constant curvature target spaces, but somewhat more challenging in general compactification manifolds.

Another proposal is to associate the inflaton with the orientation, rather than the center-of-mass position, of
a D-brane \cite{Angles1,Angles2}.

Some other interesting ideas, similar in spirit to those presented here, include \cite{HardToCategorize}.

\subsection{Warped D-brane inflation}
\subsubsection{Slow roll}

The Coulomb interaction between a D3-brane and an anti-D3-brane in a compact region of flat six-dimensional
space is generically too steep for inflation.  The proposal of \cite{KKLMMT} is that a brane-antibrane pair
separated along a warped region, such as the radial direction of a warped deformed conifold, enjoys a much
flatter Coulomb interaction.  Geometries of this sort are well-understood: they are well-approximated by
non-compact solutions, they fit naturally into the framework of type IIB moduli stabilization with fluxes
\cite{GKP}, and a useful everywhere-smooth solution is available, thanks to the work of Klebanov and Strassler
\cite{KS}.  Moreover, strongly warped compactifications allow considerable freedom in choosing energy scales. In
particular, it is worth remarking again that as discussed in \S3.1, the warped geometry has the added bonus of
redshifting down cosmic string tensions, making them a viable and interesting source of potential contact with
observations \cite{DamourVilenkin,Sarangi:2002yt,Copeland:2004iv}.

The warped brane inflation scenario \cite{KKLMMT} makes use of these important technical advantages, but in the
end is still constrained by the eta problem: a brane-antibrane pair separated along a generic warped throat does
not give rise to prolonged inflation \cite{KKLMMT}.\footnote{Much subsequent work has explored the phenomenology
of this scenario (see e.g. \cite{FurtherWarped}), and the important question of reheating was considered in
\cite{WarpedReheating}.  Another proposal for D-brane inflation in a warped throat is \cite{GiantInflaton}.}

A wide variety of solutions to this `inflaton mass problem' have been proposed.  In a compactification with a
suitable $\mathbb{Z}_{2}$ symmetry exchanging two throats, the problem can be absent \cite{Sandip}.
Alternatively, as we will review in detail in \S\ref{subsubsection:DBI}, the DBI kinetic term of a D3-brane
leads to a qualitatively new kind of inflation that does not require a nearly-flat potential, and in some
circumstances this mechanism governs the motion of a D3-brane in a warped throat.  Another proposal was that a
different sort of dynamical effect could flatten the inflaton potential \cite{Cline:2005ty}; however, this
effect relied on adjusting a parameter that was later found to be zero \cite{BDKMMM,Burgess}.

Another proposed solution is to incorporate further corrections to the inflaton potential which might, in
fine-tuned cases, cancel the problematic inflaton mass term \cite{KKLMMT}.  Such corrections are automatically
present: the nonperturbative superpotential of the KKLT scenario necessarily depends on the D3-brane position
\cite{Ganor}, and this gives rise to an effective interaction between the D3-brane and the wrapped branes
responsible for the nonperturbative superpotential \cite{BHK,GM,BDKMMM}.

Without detailed information about the functional form of this contribution to the potential, one could only
conjecture that in a certain fraction of cases, the corrected inflaton potential is sufficiently flat.  However,
this correction was determined in full, for toroidal orientifold cases, via an open string one-loop computation
in \cite{BHK}.  Subsequent work in \cite{GM} proposed that this computation could also be performed in the
closed string channel, and this was accomplished for warped throat backgrounds in \cite{BDKMMM}.  Equipped with
this result, one could ask whether the contributions to the inflaton potential indeed cancel for
specially-chosen ranges of the microscopic parameters. As shown in \cite{Burgess}, this is impossible if the
wrapped branes are described by the simple Ouyang embedding \cite{Ouyang}.  However, it was shown in
\cite{BDKMS,BDKM} and in \cite{KrausePajer} that if the wrapped branes are described by the simple and symmetric Kuperstein embedding
\cite{Kuperstein}, the potential can contain an approximate inflection point at some distance from the tip of
the throat.  Inflation can occur around this inflection point.

The resulting scenario is unusually concrete and explicit.  Even so, it will still be necessary to determine
whether the gluing of the throat region into a compact bulk causes any distortions of the supergravity
background that might correct the inflaton potential.  In addition, even if such effects are absent, the
construction is somewhat delicate: the inflection point is present only for a restricted range of parameters,
and the predictions of the scenario, including whether the scalar spectrum is red or blue, depend on fine
details of the potential.  This is an imperfect state of affairs, and it would be worthwhile to develop a more
robust and natural D-brane inflation scenario that can be realized in a concrete compactification.

\subsubsection{DBI}
\label{subsubsection:DBI}

As discussed in \S2, DBI inflation is a mechanism for slowing the motion of
scalar fields via the action (\ref{DBIEFT}), whose dynamics enforces
\be {f(\phi) \dot\phi^2} \le 1 \label{speedlimit}\ee
regardless of the steepness of the potential.

Given the effective action (\ref{DBIEFT}), there is a viable window of parameters (functions $f(\phi)$ and
$V(\phi)$) where inflation occurs, and where the density perturbations are consistent with CMB observations.
Moreover, model-independently the density perturbations are sufficiently non-Gaussian to be testable by the
Planck satellite measurements (and are already constrained by WMAP \cite{Observations,HarvardNG}). As with
$m^2\phi^2$ chaotic inflation, one would like to understand if there is an explicit UV complete model realizing
this mechanism.



The DBI mechanism for slowing down the scalar field was originally discovered in the motion on the Coulomb
branch of the ${\cal N}=4$ $U(N)$ SYM theory at large 't Hooft coupling $\lambda = g_{YM}^2 N \gg 1$, using the
AdS/CFT correspondence. On the gravity side of the correspondence, the rolling of the scalar field $\phi$ toward
the origin of moduli space corresponds to a D3-brane probe moving in the radial direction toward the horizon of
the Poincare patch of ${\rm AdS}_5$ (and sitting at a point in the ${\rm S}^5$).  This motion is governed by the
action (\ref{DBIEFT}) with warp factor $f(\phi)=\lambda/(8\pi^3 g_s\phi^4)$ and potential $V(\phi)=-1/f(\phi)$.
The DBI action enforces the causal speed limit (\ref{speedlimit}) on the motion of the probe in the gravity-side
description. The brane moves in the $r=\phi\sqrt{(2\pi)^3 g_s {\alpha^{\prime}}^2}$ direction in the space
%
%
\be ds^2={r^2\over R^2} \eta_{\mu\nu}dx^\mu dx^\nu+\frac{R^2}{r^2}\left(dr^2+r^2 d\Omega_5^2 \right)
\label{AdS}   \ee
where $R=\sqrt{\alpha^{\prime}}\lambda^{1/4}$ is the AdS curvature radius. The brane cannot move faster than the
speed of light in this background (\ref{speedlimit}). Thus, even though the distance to the origin of moduli
space $\phi=0$ is finite (and uncorrected quantum-mechanically), in the large $N$ approximation it takes forever
to reach the origin \cite{Silverstein:2003hf}.

In the CFT language, the reason for this effect is the following.  Consider taking one eigenvalue $\phi$ of the
adjoint scalar away from the origin.  This breaks $U(N)\to U(N-1)\times U(1)$.  The modes $\chi$ charged under
the $U(1)$ (and in the ${\bf{N-1}}$ representation of $U(N-1)$) have mass proportional to $\phi$. Since this
mass decreases as $\phi\to 0$, the radiative corrections from integrating out the $\chi$ multiplets generate
higher-dimension operators suppressed by powers of $\phi$ itself (rather than being suppressed by some hard high
mass scale $\Lambda$).  The first such correction, protected by supersymmetry, is $f(\phi)(\del\phi)^4$.  Since
the field theory is strongly coupled, it is difficult to sum the series of higher order terms, but the AdS/CFT
correspondence allows us to do this, revealing the result to be the DBI action (\ref{DBIEFT}) (with
$V(\phi)=-1/f(\phi)$).

With this new mechanism for slowing scalar field motion in hand, it is natural to embed it into a gravity theory
and use $\phi$ for the inflaton.  Formally cutting this theory off, coupling it to gravity, and introducing a
more general potential $V(\phi)$ in (\ref{DBIEFT}) leads to the DBI mechanism for inflation
\cite{Silverstein:2003hf,Alishahiha:2004eh}\ whose results were reviewed in \S2.  Recent work \cite{Spalinski}\
has analyzed solutions for a wider space of functions $f(\phi)$ and $V(\phi)$.

One would like to understand if this arises from an explicit string theory model.  A first set of attempts to
achieve this has been made by taking the field theory mechanism discussed above, and implementing the cutoff and
the coupling to four-dimensional gravity by embedding the throat (\ref{AdS}) in a compact manifold, matching at
some scale $\phi_{UV}$. Two regimes have been explored in more detail in the literature (the so-called ``UV
model" with potential of the form $V(\phi)=m^2\phi^2$ and Chen's ``IR model" \cite{ChenDBI}\ -- a small-field
hilltop version of DBI inflation with potential of the form $V(\phi)=V_0-m^2\phi^2$).  Of course these
two specific shapes for $V(\phi)$ are not particularly important, though they suffice to illustrate some of the
relevant features. The $m^2\phi^2$ model, which entails a Planck-scale field range and detectable gravitational
waves, cannot arise from {\it three}-branes in a Calabi-Yau throat as originally envisioned
\cite{FieldRange,Bean:2007hc,Lidsey:2007gq,Peiris:2007gz}.  The four-dimensional Planck mass $\Mpl$ in this
setup is given by
\be \Mpl^2= \frac{2\pi^3}{(2\pi)^7 g_s^2 {\alpha^\prime}^4}\int^{r_{UV}} R^4\, r\, dr =\frac{N}{4}\phi_{UV}^2
 \label{Planckmass}\ee
Thus, the field range $\phi_{UV}$ is parametrically small compared to the Planck mass \cite{FieldRange}.  This
result applies not just to the original $\rm{AdS_5\times S^5}$  geometry, but also to any warped throat
generated by D3-branes at the singularity of a cone over some Einstein manifold $X_5$.  However, for wrapped
higher-dimensional D$p$-branes rather than D3-branes \cite{BLS}\ in a Calabi-Yau or a more general
compactification, a Planckian or super-Planckian field range is geometrically possible.

Moreover, a generic model in the landscape has cycles supported by a wider variety of fluxes; moreover, the
generic compactification is not a simple Calabi-Yau manifold.  So this geometric constraint rules out an
interesting class of would-be models based on D3-branes in Calabi-Yau warped throats. But the general situation
is richer, and there is no known geometric constraint sufficient to rule out gravitational waves from DBI
inflation. However,  as in any form of inflation the possibility of gravitational waves seems to require
functional fine-tuning of the effective action.

Another challenge in obtaining DBI inflation from D-branes in a string compactification is the requirement that
the effects producing the potential $V(\phi)$ not backreact on the geometry probed by the brane, encoded in
$f(\phi)$. This affects both large-field and small-field models.  More precisely, the requirement is that once
all the leading effects are included, the entire action (\ref{DBIEFT}) is known to the required accuracy in a
solution with small generalized slow roll parameters (\ref{small}).  This is nontrivial to arrange, for the
following reason.\footnote{This argument is a four-dimensional effective field theory description of a comment
of J. Maldacena.}

Start with the field theory coupled to gravity and to a hidden sector which breaks SUSY. Given the SUSY
breaking, one generically generates a term $m^2\phi^2$. For example, in gravity mediation, $m\sim F/\Mpl$ where
$F$ is the SUSY-breaking F-term in the hidden sector.  Now, generically the same effect would lift also the
lightest modes in the throat, so the KK masses $m_{KK}\sim (1/R)(r/R)$ must satisfy $m_{KK}> m$, where $r$ is the
radial coordinate. So the smallest value of $r$, let's call it $r_{IR}$, is $r_{IR}=mR^2$, which means the
smallest value of the canonical field $\phi$ is given by
\be \phi_{IR}^2=\frac{N}{2\pi^2}m^2 \label{IRphi}\ee
This closes off the IR end of the throat at the scale $\phi_{IR}$.

Now, combine this with the requirement for inflation that the potential energy dominate over the kinetic energy:
\be V(\phi)>f^{-1}\gamma=\frac{2\pi^2}{N}\phi^{4}\gamma \label{VbeatsT}\ee
For $V(\phi)=m^2\phi^2$, this implies $\phi^2<\frac{N}{2\pi^2\gamma}m^2$.  But this is lower than the IR cutoff
$\phi_{IR}$ just estimated in (\ref{IRphi}).  One approach to evade this might be to consider not a universal
gravity-mediated contribution generating $V(\phi)$, but something analogous to gauge mediation which couples
preferentially to a brane probe.  This might require a more elaborate setup for which the charge carried by the
brane is different from the flux supporting the background AdS-like geometry.

Finally, in the case of a throat supported entirely by D3-brane charge $N$, the effective action for DBI
inflation, when matched to data, requires an enormous value for this quantum number \cite{Alishahiha:2004eh}.
However, the relations between flux/brane quantum numbers and the parameters in the effective theory are very
model-dependent. It is possible to obtain a much smaller set of input flux quantum numbers to obtain the same
low energy effective DBI action governing wrapped branes probing other compactification geometries \cite{BLS}.
Work continues to combine such a choice of branes and geometry with a solution to the other problem of
backreaction, and with an appropriate field range; recent concrete steps in this direction appeared in
\cite{BLS}\ discussed above, \cite{Spinflation}\ (which includes angular motion) and \cite{ThomasWard}\ (which
includes effects of multiple branes in DBI inflation). Once this is accomplished, analyses such as
\cite{ShiuUnderwood}\ could tie the observable parameters to some data of the compactification.

In general, in DBI as well as in slow roll inflation, there is a (roughly) one percent fine-tuning expected in
order to ensure that the generalized slow roll parameters (\ref{small}) are sufficiently small.  The DBI action
arises in a wide variety of circumstances in string theory, and it is an interesting open problem to exhibit one
in which (\ref{DBIEFT}) arises explicitly.

More generally still, DBI inflation is likely to be one of a family of models with the shared feature that the
kinetic terms are corrected by a series in $\dot\phi^2/(\phi^n M_*^{4-n})$, for some $n>1$, generated by
integrating out modes that become light as $\phi\to 0$ (here $M_*$ is a hard UV mass scale such as the KK,
string or Planck mass). It is in some sense a fortunate accident that in the case of a brane probe we know how
to sum up these effects to obtain the DBI action, but more generally {\it some} function $P(X,\phi)$
(\ref{general}) pertains in any situation of this kind.  If all terms are of the same order in the resulting
solutions, again the dynamics would tie the field velocity $\dot\phi$ to the field $\phi$ itself, leading to a
similar mechanism for slowing the scalar field.

It would be extremely interesting to understand what a generic model of inflation looks like, in terms of
$P(X,\phi)$.  A primary difficulty is defining an appropriate, and universally acceptable, notion of
genericity.\footnote{It is certainly {\it{not}} suitable for this purpose to determine what is generic among
proposed inflationary models, since humans are traditionally poor random number generators.} Moreover, it is
plausibly the case that, even equipped with such a definition, the generic models in effective theories derived
from string theory are not entirely generic in quantum field theory.  As we have explained, this is one of the
routes by which string inflation, taken as an entire framework, could eventually be predictive.

The DBI form of the action for D-branes also leads to qualitatively interesting effects on tunneling
\cite{Brown:2007ce}\ and, as argued very recently, on the cosmic string spectrum \cite{Sarangi:2007mj}.  The
question of ``assisted inflation" in this context has been studied in \cite{DBIassist}.

\subsection{Inflation in M-theory}

An interesting proposal closely analogous to D-brane inflation is M5-brane inflation, in which an M5-brane
wrapping a curve in a Calabi-Yau space moves along the interval of a Ho\v{r}ava-Witten compactification.  In the
scenario of \cite{SingleM5}, a single M5-brane very close to one of the walls moves into it.  The alternative
idea of `assisted M5-brane inflation' \cite{AssistedM5} is that a collection of many M5-branes, distributed
along the interval of the compactification, can provide a more promising inflationary potential than that of a
single M5-brane.  Inflation occurs as the M5-branes disperse and move toward the boundary walls.  In both cases,
it would be important to understand whether there is a mechanism for avoiding the eta problem: the structure of
the \k\ potential for the M5-brane position suggests that this difficulty will generically be present, as it is
in D-brane inflation scenarios.  A noteworthy feature of assisted M5-brane inflation is that, for a sufficiently
large number of M5-branes and a suitably long interval, the tensor signal could be comparatively large
\cite{Axel}.  It would be interesting to know whether these conditions can be satisfied in a consistent
compactification in which the eta problem can be addressed at the same time.

\subsection{Moduli inflation}

\subsubsection{Inflation with \k\ moduli}

Among the better-studied examples of moduli inflation are those that involve motion of the K\"ahler moduli in a
flux compactification of type IIB string theory on a Calabi-Yau threefold.  The \k\ moduli receive no potential
from imaginary-self-dual fluxes \cite{GKP}, and the idea is that the potential induced by nonperturbative
effects can be exponentially flat.

In \k\ moduli inflation \cite{Kahler}, the inflaton is the size of a `blowup' four-cycle in one of the
large-volume compactifications of \cite{BBCQ}.  A variety of contributions to the potential are suppressed by
the exponentially large overall volume of the compactification, so the leading-order potential turns out to be
very simple and extremely flat.  By the leading-order potential, we mean the potential derived by incorporating
the effect of fluxes and nonperturbative effects in the superpotential, and including the known
$\alpha^{\prime}$ corrections to the K\"ahler potential, \be {\cal{K}} = - 2
\Mpl^2~{\rm{log}}\Bigl({\cal{V(\phi)}} + \xi \Bigr) \ee Here ${\cal V}$ denotes the volume of the Calabi-Yau,
which is a function of the inflaton $\phi$, because by assumption $\phi$ is one of the \k\ moduli; the constant
$\xi$ is proportional to the Euler number of the Calabi-Yau.

An important unresolved question in this scenario is whether there exist further corrections to the \k\
potential, for example involving terms polynomial in the inflaton, \be {\cal{K}} = - 2
\Mpl^2~{\rm{log}}\Bigl({\cal{V(\phi)}} + \xi \Bigr) + f(\phi){\cal{V}}^{-c} \ee For suitable functions $f$ and
constants $c$, this correction curves the inflaton potential and leads to an eta problem.  As noted in
\cite{Kahler} and further explained in \cite{Roulette}, such corrections are readily conceivable, and one might
hope to compute them, perhaps along the lines of \cite{BHKKahler,LoopsOne,LoopsTwo}.

\k\ inflation as presently constructed is a small-field model, and as such is somewhat sensitive to the initial
conditions.  However, as argued in \cite{Roulette}, a considerable fraction of randomly-chosen trajectories can
give rise to substantial inflation.

In the type IIA orientifold constructions of \cite{DGKT,VZ,IW}, the complex structure moduli and \k\ moduli are
on equal footing.  A recent search for inflation in this context was unsuccessful \cite{Hertzberg}, however
there are further ingredients that can contribute to the potential energy.

\subsubsection{Inflation with axions}

Because axions enjoy a shift symmetry that is unbroken to all orders in perturbation theory, they are natural
candidates for inflaton fields \cite{Natural}.  The simplest scenario would be one in which a single axion has a
large periodicity, $f \gg \Mpl$ \cite{Natural}.  At present this has yet to be achieved in string theory, and
there are suggestive but incomplete arguments that this is impossible in general \cite{Gorbatov}.  As a result,
attempts to use string axions for inflation fall into two classes: those that use a small number of axions, and
those that use many axions at once.

One proposal \cite{CompletingNatural} is that under special circumstances, the effective decay constant of a
linear combination of two axions will be super-Planckian, allowing a plausible realization of natural inflation
in string theory.

Another scenario involving a small number of axions is (Better) Racetrack inflation
\cite{Racetrack,BetterRacetrack}, in which a linear combination of the axions associated to two \k\ moduli rolls
off of a saddle point of the potential.  The background compactification is the KKLT-type construction of
\cite{DDF} in an orientifold of the Calabi-Yau threefold ${\mathbb{P}}_{1,1,1,6,9}$.  Because the moduli
stabilization is comparatively well-understood, this is a rather explicit construction. One difficulty, however,
is that the parameters chosen to achieve inflation in \cite{BetterRacetrack} are somewhat strained, as noted
originally in \cite{BetterRacetrack}: in particular, the rank of one of the gauge groups whose strong gauge
dynamics contributes to the nonperturbative superpotential was taken to be $258$, which would have to correspond
to a stack of $258$ D7-branes.  It is not obvious that this can be achieved in the construction of \cite{DDF},
and it would be worthwhile to find concrete compactifications, including the data of D7-brane stacks, that
fulfill the promise of \cite{BetterRacetrack} (see also \cite{Amanda}).

Another approach \cite{Nflation} is to consider the simultaneous displacement of $ N \gg 1$ axions from their
minima, and identify an effective inflaton with the Pythagorean sum of the individual displacements.  For
sufficiently large $N$, the resulting effective inflaton has a super-Planckian range.  If the constituent axions
all had exactly the same mass and same initial displacement, this model would be indistinguishable from Linde's
original chaotic inflation model.  However, these conditions will certainly not be obeyed exactly; one therefore
needs to know the axion mass spectrum in any given construction, and the observational predictions for a model
with this spectrum.  In \cite{RMTNflation} these problems were solved using random matrix theory, and it was
shown that under very plausible assumptions, the spectrum of axion masses-squared in any KKLT compactification
is described by the Mar\v{c}enko-Pastur law.  The single parameter characterizing the shape of this distribution
turns out to be a ratio of Hodge numbers; the complicated and virtually uncomputable data associated with the
details of the compactification becomes irrelevant in the large $N$ limit \cite{RMTNflation}.  The implication
is that, although N-flation will not precisely coincide with single-field chaotic inflation, it will generically
be only slightly -- but measurably -- different: in particular, the scalar spectrum is slightly more red in
N-flation than in single-field $m^2\phi^2$ inflation.  For further investigations of the phenomenology of this
scenario, see \cite{KimLiddleOne,Piao,KimLiddleTwo,Battefeld,KimLiddleThree}.

There are at least two important concerns about N-flation.  The first is that the necessary compactifications
are near the boundary of validity of the $g_s$ and $\alpha^{\prime}$ expansions, and pack a tremendous `amount
of topology' into a small volume.  In the four-dimensional theory, this manifests as a renormalization of the
Newton constant, which may be thought of as being enhanced by the large number of species of particles
circulating in loops.  More precisely, one can determine the leading term in this renormalization from an
${\alpha^{\prime}}^3$ correction in ten dimensions.  The problem is that higher corrections in the
$\alpha^{\prime}$ and $g_s$ expansions are not known at present, and could potentially spoil the success of the
scenario.

The second concern \cite{RenataReview} is that the real-part K\"ahler moduli, which control the volumes of
four-cycles, will not find and remain in their minima as their axion partners roll towards the origin.  This
could give rise to important corrections to the potential, and in particular could negate the advantage
resulting from the axions' shift symmetry.

The constraints from reheating in N-flation were considered in \cite{Green:2007gs}.  A general problem of closed
string inflation that is particularly acute in N-flation is that in the presence of a large number of hidden
sectors, it is typically difficult to ensure that the energy of the inflaton ends up primarily in the visible
sector. In certain D-brane inflation models this problem is alleviated, because {\it{both}} the inflaton and the
visible sector are well-localized in the compact space, and if they coincide at the end of inflation, strong
reheating of the visible sector is then nearly automatic.

\subsubsection{Inflation with complex structure moduli}

At present there are very few explicit scenarios involving motion of complex structure moduli.  One reason is
that, in the type IIB flux compactifications of \cite{GKP}, the complex structure moduli receive mass from the
flux background, whereas the \k\ moduli and their axions do not; hence, the latter sector is arguably a more
promising place to look for a relatively flat potential.  However, it would be most interesting to understand
the prospects for inflation with complex structure moduli, presumably in another class of compactifications.

\subsection{Landscape inflation}

Recent progress in understanding the potential on the configuration space of string compactifications, i.e. the
`string landscape', has led to a variety of original ideas for inflation.  These proposals typically do not
involve direct specification of an inflaton candidate in a concrete setup, but rather appeal to the general
properties of the landscape, such as its high dimensionality and the large number of saddle points.

In New Old inflation \cite{NewOldOne,NewOldTwo}\ (also known as Locked inflation), a scalar field oscillates
repeatedly past a saddle point, and does not roll to the true vacuum -- ending inflation -- until the period of
its oscillations becomes small compared to the timescale set by the tachyonic mass of the saddle.  Given a large
initial amplitude and a small tachyonic mass, this leads to prolonged inflation.  However, the parameter space
is strongly constrained by limits on formation of primordial black holes \cite{LockedOne}, and it has been
argued that a long period of locked inflation must end through parametric resonance \cite{LockedTwo}.

It has also been suggested \cite{Multiple} that two or more stages of inflation, at widely-separated scales,
might involve less fine-tuning than a single period of inflation, and that such a history is plausible in the
complicated potentials found in string theory.

Another scenario inspired by the idea of a complicated landscape of vacua is Chain inflation
\cite{ChainOne,ChainTwo,ChainThree,ChainFour}.

\subsection{Other ideas}

String theory, and associated patterns of thought, have led to a diverse array of cosmological scenarios, some
quite different from the inflationary models we have discussed here.  Limitations of space, and of expertise,
prevent us from covering the cyclic and ekpyrotic scenarios, mirage cosmology, cosmology of string and brane
gases, and the Pre-big bang scenario, to name a few of the proposals most closely linked to string theory.  For
treatment of these subjects we refer the reader to
\cite{CyclicReviewOne,CyclicReviewTwo,AndreiEkpyrotic,Mirage,ScottReview,PBB} and the references therein.

\section{Dynamics of Moduli}

Let us mention a few other roles that scalar fields in field theory and string theory can play in cosmology.
First, the dynamics of moduli is relevant for the problem of vacuum selection and the question of initial
conditions for inflation. More generally, rolling scalar fields can drive non-inflationary cosmological phases,
and by characterizing the possibilities in this class, one can understand the scope of cosmological behaviors
that are allowed by string theory, as a UV completion of gravity.

\subsection{Effects of light species at weak and strong coupling}

In a Lagrangian with multiple scalar fields, scalar-scalar interactions are widespread. For example, two scalars
$\phi,\chi$ often interact via the quartic coupling
%
\be \Delta {\cal L} = g^2 \phi^2\chi^2 \label{phichi}\ee
Consider a background in which $\phi$ is rolling through its configuration space.  The interaction
(\ref{phichi}) leads to a $\phi$-dependent mass for $\chi$.  That is, the Hamiltonian for $\chi$ is
time-dependent in a background solution in which $\phi$ is rolling.  This leads to a backreaction on the
solution for $\phi$ itself. A similar effect holds for $\chi$ particles of other spins; a canonical situation
with vector $\chi$'s arises when $\phi$ is a Higgs field for the corresponding gauge symmetry.

There are two basic ways in which this happens: (i) virtual $\chi$ quanta circulating in loops renormalize the
effective action for $\phi$ (as discussed in regard to DBI inflation in \S2 and \S4.2.2) and (ii) the
time-dependent masses and couplings lead to the production of on-shell $\chi$ quanta
\cite{Traschen:1990sw,KLSpreheating,KLLMMS,Watson:2004aq,Greene:2007sa}. In the latter case, the energy density
of the resulting gas of $\chi$ particles is minimized at the point $\phi=0$ where they are massless, which leads
to a force on $\phi$ drawing it toward this point.

At strong 't Hooft coupling in AdS/CFT systems, (i) is a controlled process and dominates over (ii).  At weak
coupling, (ii) is a controlled process that dominates over (i).  In both limits (and probably also in the
intermediate regime where both occur), the effect is to trap $\phi$ near points with extra light species $\chi$.
(See \cite{DineSymmetries}\ for other work on enhanced symmetry points and the moduli problem.)

This trapping effect in case (ii) only pertains when the kinetic energy transferred into $\chi$ particles
dominates over the potential energy.
Hence, it may play a role in any epoch in which scalar fields roll relatively freely.  In an expanding universe,
the gas of $\chi$ particles is diluted over time, weakening the trapping effect, as discussed in detail in
\cite{KLLMMS}.

The above effect occurs at the level of field theory, but also extends to situations in which the $\chi$ fields
are extended strings, in such a way that oscillator modes enhance the effect when weakly-coupled branes collide
relativistically \cite{McAllister:2004gd}. Early in the study of string cosmology it was realized that the
presence of a gas of extended objects (strings and branes) could affect the evolution in interesting ways
\cite{Brandenberger:1988aj}\ in the presence of nontrivial topology. For example, in the heterotic theory on a
torus, winding strings $\chi$ become light and enhance the gauge symmetry at the self-dual radius.  A somewhat
similar effect is relevant for closed string winding modes near spacelike singularities, as discussed in \S2.

In their regime of applicability, these kinetic effects provide a dynamical mechanism favoring points with light
species.  It is important to note, however, that this does {\it{not}} in itself stabilize the moduli; the
potential energy dominates at late times.

Another interesting phase of string cosmology is that which results from the annihilation of unstable D-branes.
This is a process that is affected strongly by $\alpha'$ effects which have been extensively studied, and hence
provides a setting incorporating inherently string-theoretic behavior into cosmology.  For a review of this
direction, see \cite{Senreview}.

\subsection{Other cosmological phases}

String theory has a rich set of weakly coupled limits, unified by various dualities and by topology-changing and
even dimension-changing transitions (such as the recent classes of examples
\cite{Dduality,HellermanSC,Aharony:2006ra,Freedman:2005wx}\ discussed in \S2.2). As also discussed above, in
every known model of inflation, a certain amount of fine-tuning is required from the low energy point of view,
and the effective action derived from string theory does not appear to yield miraculous top-down cancellations
of the generalized slow roll parameters. In particular, in weak-coupling asymptotic regimes of scalar field
space (such as large volume ${\cal V}\gg l_s^{D-4}$ and weak string coupling $g_s \ll 1$), the kinetic terms are
canonical and the potential slopes toward zero too quickly for inflation.

The perturbative potential is exponential in the canonically-normalized scalar fields (linear combinations of
the dilaton $\Phi=\Mpl\,{\rm log} (g_s)$, the radion $\sigma\equiv \Mpl\,{\rm log}({\cal V})$, and so on). The form of this
potential leads to two interesting effects.  First, the generic weakly-coupled rolling scalar solution is not
inflation, but rather a rather simple quintessence model with $w=-2/3$ (on the boundary between acceleration and
deceleration). The leading term in the moduli potential scales like $g_s^2 (D-10)/{\cal V}$. For $D>10$ with a
Ricci-flat internal compactification there is an $\alpha'$-exact tree-level solution
\cite{Myers:1987fv,otheroldSC}\ with scale factor $a(t)\propto t$ linear in FRW proper time $t$ (and with the
dilaton linear in conformal (string frame) time $X^0\sqrt{D}\sim l_s \,{\rm log}(t/l_s)$). In this $a(t)\propto t$
phase arising generically in weakly-coupled string theory, modes neither enter nor leave the horizon, and the
question of the spectrum of density perturbations reverts to one of initial conditions, tying it to the physics
closer to the initial singularity (for recent studies of this, see
\cite{HellermanSC,Freedman:2005wx,Aharony:2006ra}).  The second intriguing feature of these weakly-coupled
backgrounds is the following. A canonically-normalized field with a negative exponential potential has a
scale-invariant power spectrum in Minkowski space, and also in any background in which its energy density is
subdominant in driving the expansion of the universe.  (See \cite{Creminelli:2007aq}\ and references therein for
a recent discussion of this.) Backgrounds with a compactification manifold of positive scalar curvature, or with
subcritical dimensionality, yield a negative exponential potential coming from string theory.

Work continues aimed at understanding the role of these more general backgrounds. Because of the genericity of
the exponential potential terms\footnote{In \cite{Itzhaki} it is proposed that suitable collections of
exponential potentials produce inflationary models which, although still fine-tuned, are arguably numerous.} and
the possibility of generating scale-invariant perturbations from them, it would be very interesting to
understand if the generic effective theory descending from string theory yields an alternative to inflation.
However, this is extremely challenging -- at present inflation fits the observations in numerous interconnected
ways, making it very difficult to envision a competitive alternative.

\section{Dark Energy in String Theory}

The discovery that the expansion of the universe is accelerating is among the most significant of recent times.
It makes untenable the old hope that some mechanism might set the cosmological constant precisely to zero, and
it demands an explanation of why the scale of dark energy is nonzero, yet so small compared to the Planck scale,
or indeed even to the \gev\ scale.

Our present understanding of string theory, albeit incomplete, meshes with a nonzero value for the cosmological
constant. There is a ``discretuum" of contributions to the moduli potential arising from choices of discrete
quantum numbers (dimensionality, topology, fluxes, brane and orientifold numbers, and so on)
\cite{Bousso:2000xa}\ which plausibly yields enough freedom to tune the cosmological constant finely, but not
precisely to zero.  This argument has two caveats:  (1) it assumes moduli stabilization, and (2) it assumes that
there are no conspiracies affecting this tuning process in the realistic window in which $\ge \rm TeV$ scale
quantum corrections would need to be tuned almost away.  Point (1) has been addressed extensively in recent and
ongoing work, while point (2) remains inaccessible to explicit analysis.

Various mechanisms of of moduli stabilization have been the subject of several useful lectures and reviews, such
as \cite{Silverstein:2004id,Grana:2005jc,PolchinskiCC,Douglas:2006es,Denef:2007pq}, so we will only give a
qualitative picture here. The basic problem is that the potential slopes to zero in weakly coupled limits such
as large radius (in geometric models) and weak string coupling (in corners of M-theory with a perturbative
string description).

If we focus on geometrical compactifications near weak-coupling limits, it is straightforward to see that
sources of positive energy in four-dimensional effective theories derived from string compactifications source
an expansion of the internal space.  This can be understood as follows.  Begin with a source of positive energy
in $D$ dimensions, including the tree-level cosmological term proportional to $D-10$, as well as additional
sources such as brane or flux configurations, and dimensionally reduce to four-dimensional Einstein frame. The
volume ${\cal{V}}$ of the internal space enters the rescaling to Einstein frame in such a way that these
higher-dimensional energy sources, when reduced to four-dimensional potentials ${\cal{U}}_i$, scale inversely
with the volume, \beq {\cal{U}}_i = \frac{c_i}{{\cal{V}}^{\alpha_i}} \eeq with $\alpha_i$ positive and $c_i$
independent of ${\cal{V}}$.  Hence, sources of positive energy in string theory tend to yield a runaway
behavior, driving the volume to larger values. The result is that a stable or metastable de Sitter vacuum is
possible only in the presence of a counterbalancing force that prevents decompactification.  Similar statements
apply to the string coupling.  In other words, the simplest way to obtain de Sitter space requires three
competing terms in the potential expanded about weak coupling:
\be {\cal U}\sim a g^2 - b g^3 +
c g^4 \dots \label{threeterms} \ee
where $g$ represents the string coupling and the (moduli-dependent) coefficients  $a,b,c$ are all positive at a
local minimum of the potential; a similar expansion obtains for inverse volumes and any other runaway moduli in
the problem \cite{DineSeiberg}. A little algebra reveals that for a de Sitter solution, one needs sufficient
freedom to tune the coefficients at the minimum to obtain $0<(4ac/b^2)-1<1/8$. In particular, the negative term
$-b g^3$ needs to be sufficiently strong relative to the leading $a g^2$ term, and involves ingredients (such as
orientifold planes or closely-related objects in all known constructions) going beyond GR or supergravity.  As
emphasized in \cite{Silverstein:2001xn}, this leads to solutions which are only {\it meta-}stable, a result that
follows rather generally from considerations of thermal effects in local de Sitter solutions
\cite{KKLT,laterlenny}.  For a recent analysis of decay rates in some constructions, see \cite{Westphal:2007xd}.

In general, it is essential to recognize that one must introduce a small or large number by hand in order to
ensure that different orders in perturbation theory (as well as nonperturbative effects) can compete with each
other to fix the runaway moduli.  To ensure a small value of the vacuum energy at the minimum, much further
tuning would be required.

In the last several years this problem has been the focus of intense efforts, and a range of mechanisms have
been proposed at various scales of supersymmetry breaking
\cite{Maloney:2002rr,KKLT,Saltman:2004jh}.\footnote{These works followed early works on flux compactification,
such as \cite{Becker:2001pm,Dasgupta:1999ss,Gukov:1999ya,GKP,Silverstein:2001xn}.} In particular, the proposal
of \cite{GKP,KKLT}\ (and subsequent work) builds up from a Ricci-flat internal space -- a special choice which
is compatible with traditionally natural solutions to the hierarchy problem: low energy supersymmetry combined
with Randall-Sundrum warped extra dimensions. The others \cite{Maloney:2002rr,Saltman:2004jh}\ build from the
leading terms in the moduli potential (including the positive potential energy arising from supercriticality
and/or negative curvature), arising from a much wider range of choices of topologies and dimensionalities for
the internal space -- but they do not incorporate a traditionally natural solution to the weak hierarchy
problem.

One interesting tradeoff in
the subject is that the higher scales of SUSY breaking permit more terms in the effective action.  This
simplifies the analysis in some ways and complicates it in others.  It means on the one hand that more competing
forces are available in perturbation theory in $g$, allowing for metastabilization at this level. On the other
hand, more corrections are allowed and must be bounded in order to exhibit a well-defined approximation scheme.
In the low-energy SUSY framework of \cite{GKP,KKLT}, SUSY constrains the effective action so that one needs to
use a non-perturbative contribution, competing with a perturbative flux superpotential. On the other hand, SUSY
helps stabilize against decays and corrections, simplifying that part of the analysis. In any case, at any scale
of SUSY breaking, control is obtained via perturbation theory, by ensuring that $\alpha'$ and $g$
perturbations about the background solution are small.

As with the models of inflation discussed previously, we wish to make clear some of the open issues in this
area. A concern about the framework \cite{Maloney:2002rr}, as mentioned in \cite{Maloney:2002rr}, is the large
number of RR-sector species at high dimension $D$, though as in the critical dimension one expects Chern-Simons
couplings to lift many RR axions,\footnote{E.S. thanks S. Hellerman for this comment.}\ and loop effects in
high-dimensional field theory are suppressed by factors of $1/D!$ at large $D$ \cite{StromingerD}. An open
question about the framework \cite{GKP,KKLT}\ and also \cite{Saltman:2004jh}\ is the role of the full complement
of ${\cal R}^4$ terms. A specific quartic curvature term related to the Euler character of an F-theory fourfold
is used in the construction. This term has a topological reason to have a large coefficient, so it is reasonable
to focus first on its effects.  As reviewed in detail in \cite{Douglas:2006es}, a subset of the remaining quartic curvature terms has been computed in a number of important works, and the result agrees with the expectation that the Euler character term dominates.  Even so, it would be interesting to understand explicitly whether other ${\cal R}^4$ terms might compete in some circumstances, and if so how they affect the solution.

The discussion of the landscape also led to some spurious objections to the model-building efforts reviewed
here. Some physicists objected to the use of RR fields.  RR fields appear in the simplest flux compactifications
-- Freund-Rubin models -- which admit an exact formulation based on AdS/CFT, and their use fits into standard
effective field theory. Similar comments apply to the objection to models based on gaugino condensation and
other non-perturbative effects; the effects included in \cite{KKLT}\ are standard contributions in the low
energy effective theory which forms the basis of a well-defined approximation scheme in the models. Some
physicists objected to the supercritical examples based on the claim that these are not part of M theory. This
is incorrect.  For recent analyses exhibiting the dynamical connection between critical and non-critical corners
of string theory see \cite{Dduality,HellermanSC}\ (reviewed above in \S2).\footnote{It has been known since the
early years of string theory that tachyon condensation lowers the effective central charge, the only invariant
consistency condition being that the total central charge (matter plus ghost) vanishes on the worldsheet.} In
particular, the ten-dimensional superstring on an expanding compact negatively curved space has a supercritical
effective central charge; Hubble expansion suffices to exhibit the transition from supercritical to critical
regimes of string theory.  Moreover, tachyons tuned to turn on along a lightlike direction provide an
$\alpha'$-exact method for exhibiting dynamical connections between highly supercritical and critical limits of
string theory.  Nonetheless, although some of the objections have been incorrect, it is fair to say that all
models in the landscape are somewhat complicated, and it is important to bear in mind the possibility -- however
unlikely -- that some new top-down consistency criterion will eliminate some or all of the proposed solutions.

It will be very interesting to learn from the LHC whether low-energy supersymmetry governs the electroweak
scale; the landscape of string vacua has reinvigorated discussion of the open question of how similar (or not)
are the problems of explaining the electroweak and cosmological constant scales (see for example \cite{split}).
In any case, although it is now clear that the requisite competing forces are available in reasonably
well-understood compactifications, nothing close to a complete picture of the string landscape presently exists.
It is interesting to develop completely explicit examples of ${\rm{dS}}_4$ in string theory, as has been
achieved for ${\rm{AdS}}_4$ flux compactifications in \cite{DDF,Conlon:2004ds,Denef:2005mm}\ and especially
simply in \cite{DGKT}; there are new results in this direction (e.g. \cite{ESdS}).  However, explicit,
controlled weakly coupled models of de Sitter along these lines would not be realistic, since that would require
explicit fine-tuned cancellation of loops of Standard Model particles.  For this and other reasons, another
important direction, which we will not review here, is the quest for concrete holographic duals of landscape
models.

Of course it would be spectacular to find a first-principles prediction of the scale of dark energy rather than
accommodating it as a selection effect \cite{Weinberg:1987dv}, but this is extremely challenging
\cite{Weinberg:2000yb}. At the time of this writing, no compelling proposal exists, despite some interesting
attempts.\footnote{One related point to emphasize is that it appears much simpler to obtain a local metastable
minimum for the moduli, rather than further fine-tuning to obtain quintessence fields or other very light
scalars; however, interesting and testable scenarios such as \cite{Khoury:2003rn}\ have been explored.}
Moreover, the success of inflationary cosmology as well as the presence of a discretuum of string vacua appears to point
instead toward a global cosmology containing a wide variety of regions with different local cosmological
parameters.\footnote{There are ideas for seeking observational evidence for a tunneling transition preceding
our local cosmology \cite{Freivogel:2005vv}.}

\section{Conclusions}

The outlook for string cosmology is rather bright: at least for the near future, experimental and theoretical
progress are likely to be rapid.  The coming generation of cosmological experiments will probably rule out the
great majority of string inflation scenarios, as well as most of their alternatives, allowing a more efficient
focus on the surviving models.  Steady progress in string theory should ensure that cosmological model-building
ten years from now will be substantially more realistic and rigorous than it is now. We also anticipate progress
towards a more comprehensive theory of initial conditions and spacelike singularities.

One of the lessons of the subject so far, it seems to us, is the presence of serendipitous connections between
formal developments and observational opportunities. Focusing on mechanisms (such as moduli stabilization
effects and warping, or the speed limit on brane probes) led to inflation scenarios with detailed observational
predictions (such as viable cosmic string tensions, or non-Gaussian corrections to the CMB) in a way that would
have been difficult to predict from the start. Moreover, spinoffs of string theory such as novel inflation
mechanisms may be considered independently within effective field theory, whether or not they are realized
concretely or generically in string theory--supersymmetry being a famous example of this within particle
physics. Of course these connections are only truly exciting if observed in experiments, but they sometimes
yield important conceptual lessons in the meantime.

Despite these promising signs, it remains to be seen whether this endeavor will lead to genuine contact between
experiment and Planck-scale physics. In many scenarios, inflation is described by a well-controlled, albeit
fine-tuned, effective field theory Lagrangian, and inflation lasts long enough to obscure all evidence of a
pre-inflationary stage. If we live in such a universe, cosmological observations can, at best, teach us about
the nature of the inflaton, but will provide few clues about more fundamental physics, except perhaps through
the enduring mystery of dark energy.

\section*{Acknowledgments}

We are grateful to all our collaborators on this subject and to many others for useful discussions during the
various developments reviewed here. We particularly thank D. Baumann and S. Kachru for many helpful discussions
during the preparation of this review and for comments on a draft. We also thank the KITP Santa Barbara for
hospitality during the initiation of this work. The research of E.S. is currently supported by NSF grant
PHY-0244728, by the DOE under contract DE-AC03-76SF00515, and by BSF and FQXi grants.

\newpage
\begingroup\raggedright\endgroup


\begin{thebibliography}{10}

\baselineskip=14.5pt

\bibitem{Supernova}
  A.~G.~Riess {\it et al.}  [Supernova Search Team Collaboration],
  Astron.\ J.\  {\bf 116}, 1009 (1998)
  [arXiv:astro-ph/9805201];

  S.~Perlmutter {\it et al.}  [Supernova Cosmology Project Collaboration],
  Astrophys.\ J.\  {\bf 517}, 565 (1999)
  [arXiv:astro-ph/9812133].


\bibitem{Observations}
  D.~N.~Spergel {\it et al.}  [WMAP Collaboration],
  Astrophys.\ J.\ Suppl.\  {\bf 148}, 175 (2003)  [arXiv:astro-ph/0302209];

  H.~V.~Peiris {\it et al.},
  Astrophys.\ J.\ Suppl.\  {\bf 148}, 213 (2003)
  [arXiv:astro-ph/0302225];

  D.~N.~Spergel {\it et al.},
  [arXiv:astro-ph/0603449].


\bibitem{Inflation}
  A.~H.~Guth,
  Phys.\ Rev.\ D {\bf 23}, 347 (1981);

  A.~D.~Linde,
  Phys.\ Lett.\ B {\bf 108}, 389 (1982);

  A.~Albrecht and P.~J.~Steinhardt,
  Phys.\ Rev.\ Lett.\  {\bf 48}, 1220 (1982).


\bibitem{Mukhanov:2005sc}
  V.~Mukhanov,
  ``Physical foundations of cosmology,''
{\it  Cambridge, UK: Univ. Pr. (2005) 421 p}



\bibitem{Weinberg:1987dv}
  S.~Weinberg,
  Phys.\ Rev.\ Lett.\  {\bf 59}, 2607 (1987).



\bibitem{Becker:2001pm}
  K.~Becker and M.~Becker,
  JHEP {\bf 0107}, 038 (2001)
  [arXiv:hep-th/0107044].

\bibitem{Dasgupta:1999ss}
  K.~Dasgupta, G.~Rajesh and S.~Sethi,
  JHEP {\bf 9908}, 023 (1999)
  [arXiv:hep-th/9908088].

\bibitem{Gukov:1999ya}
  S.~Gukov, C.~Vafa and E.~Witten,
  Nucl.\ Phys.\  B {\bf 584}, 69 (2000)
  [Erratum-ibid.\  B {\bf 608}, 477 (2001)]
  [arXiv:hep-th/9906070].


\bibitem{GKP}
  S.~B.~Giddings, S.~Kachru and J.~Polchinski,
  Phys.\ Rev.\  D {\bf 66}, 106006 (2002)
  [arXiv:hep-th/0105097].

\bibitem{Silverstein:2001xn}
  E.~Silverstein,
  arXiv:hep-th/0106209.

\bibitem{Maloney:2002rr}
  A.~Maloney, E.~Silverstein and A.~Strominger,
  arXiv:hep-th/0205316.

\bibitem{KKLT}
  S.~Kachru, R.~Kallosh, A.~Linde and S.~P.~Trivedi,
  Phys.\ Rev.\  D {\bf 68}, 046005 (2003)
  [arXiv:hep-th/0301240].



\bibitem{Frey:2003tf}
  A.~R.~Frey,
  arXiv:hep-th/0308156.

\bibitem{Silverstein:2004id}
  E.~Silverstein,
  arXiv:hep-th/0405068.


\bibitem{Grana:2005jc}
  M.~Grana,
  Phys.\ Rept.\  {\bf 423}, 91 (2006)
  [arXiv:hep-th/0509003].

\bibitem{PolchinskiCC}
  J.~Polchinski,
  arXiv:hep-th/0603249.


\bibitem{Douglas:2006es}
  M.~R.~Douglas and S.~Kachru,
  arXiv:hep-th/0610102.



\bibitem{Denef:2007pq}
  F.~Denef, M.~R.~Douglas and S.~Kachru,
  arXiv:hep-th/0701050.

\bibitem{Boussoreview}
  R.~Bousso,
  arXiv:0708.4231 [hep-th].





\bibitem{Bousso:2000xa}
  R.~Bousso and J.~Polchinski,
  JHEP {\bf 0006}, 006 (2000)
  [arXiv:hep-th/0004134].

\bibitem{FMRSW}
  J.~L.~Feng, J.~March-Russell, S.~Sethi and F.~Wilczek,
  Nucl.\ Phys.\  B {\bf 602}, 307 (2001)
  [arXiv:hep-th/0005276].








\bibitem{reviews}

  S.~H.~Henry Tye,
  arXiv:hep-th/0610221.

J.~M.~Cline,
  arXiv:hep-th/0612129.

R.~Kallosh,
  arXiv:hep-th/0702059.

C.~P.~Burgess,
  arXiv:0708.2865 [hep-th].




\bibitem{Borde:2001nh}
S.~W.~Hawking and R.~Penrose,
  Proc.\ Roy.\ Soc.\ Lond.\  A {\bf 314}, 529 (1970).

A.~Borde, A.~H.~Guth and A.~Vilenkin,
  Phys.\ Rev.\ Lett.\  {\bf 90}, 151301 (2003)
  [arXiv:gr-qc/0110012].



\bibitem{StromingerVafa}
  A.~Strominger and C.~Vafa,
  Phys.\ Lett.\  B {\bf 379}, 99 (1996)
  [arXiv:hep-th/9601029].


\bibitem{Garytalk}
G.~T.~Horowitz, 
www.damtp.cam.ac.uk/estg06/talks/horowitz/horowitz.ppt

\bibitem{Berkooz:2007nm}
  M.~Berkooz and D.~Reichmann,
  arXiv:0705.2146 [hep-th].

\bibitem{Craps:2006yb}
  B.~Craps,
  Class.\ Quant.\ Grav.\  {\bf 23}, S849 (2006)
  [arXiv:hep-th/0605199].

\bibitem{Cornalba:2003kd}
  L.~Cornalba and M.~S.~Costa,
  Fortsch.\ Phys.\  {\bf 52}, 145 (2004)
  [arXiv:hep-th/0310099].



\bibitem{Damour:2002et}
  T.~Banks, W.~Fischler and L.~Motl,
  JHEP {\bf 9901}, 019 (1999)
  [arXiv:hep-th/9811194].

  T.~Damour, M.~Henneaux and H.~Nicolai,
Class.\ Quant.\ Grav.\  {\bf 20}, R145 (2003)
  [arXiv:hep-th/0212256].

  J.~Brown, O.~J.~Ganor and C.~Helfgott,
  JHEP {\bf 0408}, 063 (2004)
  [arXiv:hep-th/0401053].

  T.~Damour, A.~Hanany, M.~Henneaux, A.~Kleinschmidt and H.~Nicolai,
  Gen.\ Rel.\ Grav.\  {\bf 38}, 1507 (2006)
  [arXiv:hep-th/0604143].

  M.~Henneaux, M.~Leston, D.~Persson and P.~Spindel,
  cosmology,''
  JHEP {\bf 0610}, 021 (2006)
  [arXiv:hep-th/0606123].

  F.~Englert, L.~Houart, A.~Kleinschmidt, H.~Nicolai and N.~Tabti,
  JHEP {\bf 0705}, 065 (2007)
  [arXiv:hep-th/0703285].

  T.~Damour,
  arXiv:0704.1457 [gr-qc].






\bibitem{Belinsky:1970ew}
  V.~A.~Belinsky, I.~M.~Khalatnikov and E.~M.~Lifshitz,
  Adv.\ Phys.\  {\bf 19}, 525 (1970).

\bibitem{Garfinkle:2003bb}
  D.~Garfinkle,
  Phys.\ Rev.\ Lett.\  {\bf 93}, 161101 (2004)
  [arXiv:gr-qc/0312117].



\bibitem{holography}
G.~L.~Alberghi, D.~A.~Lowe and M.~Trodden,
  JHEP {\bf 9907}, 020 (1999)
  [arXiv:hep-th/9906047].

  A.~Strominger,
  JHEP {\bf 0111}, 049 (2001)
  [arXiv:hep-th/0110087].

  T.~Banks and W.~Fischler,
  arXiv:hep-th/0111142.


  M.~Alishahiha, A.~Karch, E.~Silverstein and D.~Tong,
  AIP Conf.\ Proc.\  {\bf 743}, 393 (2005)
  [arXiv:hep-th/0407125].

  M.~Alishahiha, A.~Karch and E.~Silverstein,
  JHEP {\bf 0506}, 028 (2005)
  [arXiv:hep-th/0504056].

  B.~Freivogel, V.~E.~Hubeny, A.~Maloney, R.~Myers, M.~Rangamani and S.~Shenker,
  JHEP {\bf 0603}, 007 (2006)
  [arXiv:hep-th/0510046].

  B.~Freivogel, Y.~Sekino, L.~Susskind and C.~P.~Yeh,
  Phys.\ Rev.\  D {\bf 74}, 086003 (2006)
  [arXiv:hep-th/0606204].

  N.~Arkani-Hamed, S.~Dubovsky, A.~Nicolis, E.~Trincherini and G.~Villadoro,
  JHEP {\bf 0705}, 055 (2007)
  [arXiv:0704.1814 [hep-th]].



\bibitem{Maldacena:2002vr}

J.~M.~Maldacena,
  JHEP {\bf 0305}, 013 (2003)
  [arXiv:astro-ph/0210603].

V.~Acquaviva, N.~Bartolo, S.~Matarrese and A.~Riotto,
  Nucl.\ Phys.\  B {\bf 667}, 119 (2003)
  [arXiv:astro-ph/0209156].

N.~Bartolo, E.~Komatsu, S.~Matarrese and A.~Riotto,
  Phys.\ Rept.\  {\bf 402}, 103 (2004)
  [arXiv:astro-ph/0406398].


\bibitem{measures}

  R.~Easther, E.~A.~Lim and M.~R.~Martin,
  JCAP {\bf 0603}, 016 (2006)
  [arXiv:astro-ph/0511233].

  R.~Bousso,
  Phys.\ Rev.\ Lett.\  {\bf 97}, 191302 (2006)
  [arXiv:hep-th/0605263].

  A.~Vilenkin,
  J.\ Phys.\ A  {\bf 40}, 6777 (2007)
  [arXiv:hep-th/0609193].

  A.~Linde,
  JCAP {\bf 0706}, 017 (2007)
  [arXiv:0705.1160 [hep-th]].

  T.~Clifton, S.~Shenker and N.~Sivanandam,
  arXiv:0706.3201 [hep-th].

\bibitem{tunneling}
  D.~Podolsky and K.~Enqvist,
  arXiv:0704.0144 [hep-th].

S.~H.~Tye,
  arXiv:0708.4374 [hep-th].




\bibitem{Thurston}
W. Thurston,
Bull. Amer. Math. Soc.(N.S.) 6 (1982), 357–381.

\bibitem{Milnor}
J. Milnor,
J. Diff. Geom 2, 1968 p. 1-7.

\bibitem{BondHyperbolic}
  J.~R.~Bond, D.~Pogosian and T.~Souradeep,
  Phys.\ Rev.\  D {\bf 62}, 043005 (2000)
  [arXiv:astro-ph/9912124].

J.~R.~Bond, D.~Pogosian and T.~Souradeep,
  Phys.\ Rev.\  D {\bf 62}, 043006 (2000)
  [arXiv:astro-ph/9912144].




\bibitem{Kutasov:1990sv}
  D.~Kutasov and N.~Seiberg,
  Nucl.\ Phys.\  B {\bf 358}, 600 (1991).



\bibitem{Dduality}
  E.~Silverstein,
  Phys.\ Rev.\  D {\bf 73}, 086004 (2006)
  [arXiv:hep-th/0510044].

  J.~McGreevy, E.~Silverstein and D.~Starr,
  Phys.\ Rev.\  D {\bf 75}, 044025 (2007)
  [arXiv:hep-th/0612121].

  D.~Green, A.~Lawrence, J.~McGreevy, D.~R.~Morrison and E.~Silverstein,
  Phys.\ Rev.\  D {\bf 76}, 066004 (2007)
  [arXiv:0705.0550 [hep-th]].


\bibitem{Aharony:2006ra}
  O.~Aharony and E.~Silverstein,
  Phys.\ Rev.\  D {\bf 75}, 046003 (2007)
  [arXiv:hep-th/0612031].


\bibitem{HellermanSC}

  S.~Hellerman,
  arXiv:hep-th/0405041.

  S.~Hellerman and X.~Liu,
  arXiv:hep-th/0409071.

  S.~Hellerman and I.~Swanson,
  arXiv:hep-th/0611317.

  S.~Hellerman and I.~Swanson,
  arXiv:hep-th/0612051.

  S.~Hellerman and I.~Swanson,
  arXiv:hep-th/0612116.

   S.~Hellerman and I.~Swanson,
  arXiv:0705.0980 [hep-th].



\bibitem{ChicagoLMS}
  M.~Berkooz, B.~Craps, D.~Kutasov and G.~Rajesh,
  JHEP {\bf 0303}, 031 (2003)
  [arXiv:hep-th/0212215].


\bibitem{Liu:2002yd}
  H.~Liu, G.~W.~Moore and N.~Seiberg,
  arXiv:gr-qc/0301001.

\bibitem{Berkooz:2004re}
  M.~Berkooz, B.~Pioline and M.~Rozali,
  JCAP {\bf 0408}, 004 (2004)
  [arXiv:hep-th/0405126].


\bibitem{Berkooz:2004yy}
  M.~Berkooz, B.~Durin, B.~Pioline and D.~Reichmann,
  JCAP {\bf 0410}, 002 (2004)
  [arXiv:hep-th/0407216].


\bibitem{Horowitz:2002mw}
  G.~T.~Horowitz and J.~Polchinski,
  Phys.\ Rev.\  D {\bf 66}, 103512 (2002)
  [arXiv:hep-th/0206228].

  A.~Lawrence,
  JHEP {\bf 0211}, 019 (2002)
  [arXiv:hep-th/0205288].

\bibitem{nappiwitten}
  C.~R.~Nappi and E.~Witten,
  Phys.\ Lett.\  B {\bf 293}, 309 (1992)
  [arXiv:hep-th/9206078].



\bibitem{McGreevy:2005ci}
M.~Fabinger and P.~Horava,
  Nucl.\ Phys.\  B {\bf 580}, 243 (2000)
  [arXiv:hep-th/0002073].

A.~Adams, J.~Polchinski and E.~Silverstein,
  JHEP {\bf 0110}, 029 (2001)
  [arXiv:hep-th/0108075].

  E.~J.~Martinec,
  arXiv:hep-th/0305148.

  M.~Headrick, S.~Minwalla and T.~Takayanagi,
  Class.\ Quant.\ Grav.\  {\bf 21}, S1539 (2004)
  [arXiv:hep-th/0405064].

  J.~L.~F.~Barbon and E.~Rabinovici,
  arXiv:hep-th/0407236.




  A.~Adams, X.~Liu, J.~McGreevy, A.~Saltman and E.~Silverstein,
  JHEP {\bf 0510}, 033 (2005)
  [arXiv:hep-th/0502021].

  J.~McGreevy and E.~Silverstein,
  JHEP {\bf 0508}, 090 (2005)
  [arXiv:hep-th/0506130].

  G.~T.~Horowitz,
  JHEP {\bf 0508}, 091 (2005)
  [arXiv:hep-th/0506166].

  G.~T.~Horowitz and E.~Silverstein,
  Phys.\ Rev.\  D {\bf 73}, 064016 (2006)
  [arXiv:hep-th/0601032].

  E.~Silverstein,
  arXiv:hep-th/0602230.

  Y.~Nakayama, S.~J.~Rey and Y.~Sugawara,
  arXiv:hep-th/0606127.

  M.~Berkooz, Z.~Komargodski and D.~Reichmann,
  arXiv:0706.0610 [hep-th].

  M.~Rangamani and S.~F.~Ross,
  arXiv:0706.0663 [hep-th].

P.~Horava and C.~A.~Keeler,
  arXiv:0709.3296 [hep-th].


\bibitem{Strominger:2003fn}
  A.~Strominger and T.~Takayanagi,
  Adv.\ Theor.\ Math.\ Phys.\  {\bf 7}, 369 (2003)
  [arXiv:hep-th/0303221].





\bibitem{Banks:1996vh}
  T.~Banks, W.~Fischler, S.~H.~Shenker and L.~Susskind,
  Phys.\ Rev.\  D {\bf 55}, 5112 (1997)
  [arXiv:hep-th/9610043].

\bibitem{Aharony:1999ti}
  O.~Aharony, S.~S.~Gubser, J.~M.~Maldacena, H.~Ooguri and Y.~Oz,
  Phys.\ Rept.\  {\bf 323}, 183 (2000)
  [arXiv:hep-th/9905111].


\bibitem{Kraus:2002iv}
  P.~Kraus, H.~Ooguri and S.~Shenker,
  Phys.\ Rev.\  D {\bf 67}, 124022 (2003)
  [arXiv:hep-th/0212277].

\bibitem{Fidkowski:2003nf}
  L.~Fidkowski, V.~Hubeny, M.~Kleban and S.~Shenker,
  JHEP {\bf 0402}, 014 (2004)
  [arXiv:hep-th/0306170].

\bibitem{Hertog:2004rz}
  T.~Hertog and G.~T.~Horowitz,
  JHEP {\bf 0407}, 073 (2004)
  [arXiv:hep-th/0406134].

\bibitem{HongAdSsing}
  G.~Festuccia and H.~Liu,
  JHEP {\bf 0604}, 044 (2006)
  [arXiv:hep-th/0506202].



\bibitem{Craps:2005wd}
  B.~Craps, S.~Sethi and E.~P.~Verlinde,
  JHEP {\bf 0510}, 005 (2005)
  [arXiv:hep-th/0506180].



\bibitem{Das:2006pw}
T.~Ishino, H.~Kodama and N.~Ohta,
 Phys.\ Lett.\  B {\bf 631} (2005) 68  [arXiv:hep-th/0509173]

T.~Ishino and N.~Ohta,
 Phys.\ Lett.\  B {\bf 638} (2006) 105  [arXiv:hep-th/0603215]

  S.~R.~Das, J.~Michelson, K.~Narayan and S.~P.~Trivedi,
  Phys.\ Rev.\  D {\bf 75}, 026002 (2007)
  [arXiv:hep-th/0610053].



\bibitem{Garriga:1999vw}
  J.~Garriga and V.~F.~Mukhanov,
  Phys.\ Lett.\  B {\bf 458}, 219 (1999)
  [arXiv:hep-th/9904176].

\bibitem{InflationEFT}
  C.~Cheung, P.~Creminelli, A.~L.~Fitzpatrick, J.~Kaplan and L.~Senatore,
  arXiv:0709.0293 [hep-th].


\bibitem{BDKM}
  D.~Baumann, A.~Dymarsky, I.~R.~Klebanov and L.~McAllister,
  arXiv:0706.0360 [hep-th].

\bibitem{Chen:2006nt}
  X.~Chen, M.~x.~Huang, S.~Kachru and G.~Shiu,
  JCAP {\bf 0701}, 002 (2007)
  [arXiv:hep-th/0605045].



\bibitem{Starobinsky}
  A.~A.~Starobinsky,
  JETP Lett.\  {\bf 30}, 682 (1979)
  [Pisma Zh.\ Eksp.\ Teor.\ Fiz.\  {\bf 30}, 719 (1979)].

\bibitem{Lyth}
  D.~H.~Lyth,
  Phys.\ Rev.\ Lett.\  {\bf 78}, 1861 (1997)
  [arXiv:hep-ph/9606387].

\bibitem{ShamitTalk}
S. Kachru, talks at Solvay Conference, November 2005; KITP String Phenomenology workshop, September 2006; and
COSMO 06, September 2006.




\bibitem{CreminelliHD}
P.~Creminelli,
  JCAP {\bf 0310} (2003) 003
  [arXiv:astro-ph/0306122].

  N.~Arkani-Hamed, P.~Creminelli, S.~Mukohyama and M.~Zaldarriaga,
  JCAP {\bf 0404} (2004) 001
  [arXiv:hep-th/0312100].



\bibitem{Silverstein:2003hf}
  E.~Silverstein and D.~Tong,
  Phys.\ Rev.\  D {\bf 70}, 103505 (2004)
  [arXiv:hep-th/0310221].


\bibitem{Alishahiha:2004eh}
  M.~Alishahiha, E.~Silverstein and D.~Tong,
  Phys.\ Rev.\  D {\bf 70}, 123505 (2004)
  [arXiv:hep-th/0404084].



\bibitem{InflationEFTconsistency}
  C.~Cheung, A.~L.~Fitzpatrick, J.~Kaplan and L.~Senatore,
  arXiv:0709.0295 [hep-th].



\bibitem{HarvardNG}
  P.~Creminelli, L.~Senatore, M.~Zaldarriaga and M.~Tegmark,
  JCAP {\bf 0703} (2007) 005
  [arXiv:astro-ph/0610600].


  P.~Creminelli, L.~Senatore and M.~Zaldarriaga,
  JCAP {\bf 0703} (2007) 019
  [arXiv:astro-ph/0606001].


  P.~Creminelli, A.~Nicolis, L.~Senatore, M.~Tegmark and M.~Zaldarriaga,
  JCAP {\bf 0605} (2006) 004
  [arXiv:astro-ph/0509029].


  D.~Babich, P.~Creminelli and M.~Zaldarriaga,
  JCAP {\bf 0408} (2004) 009
  [arXiv:astro-ph/0405356].

  P.~Creminelli and M.~Zaldarriaga,
  JCAP {\bf 0410} (2004) 006
  [arXiv:astro-ph/0407059].

\bibitem{Easther:2002xe}
  R.~Easther, B.~R.~Greene, W.~H.~Kinney and G.~Shiu,
  Phys.\ Rev.\  D {\bf 66}, 023518 (2002)
  [arXiv:hep-th/0204129].


\bibitem{Kaloper:2002uj}
  N.~Kaloper, M.~Kleban, A.~E.~Lawrence and S.~Shenker,
  Phys.\ Rev.\  D {\bf 66}, 123510 (2002)
  [arXiv:hep-th/0201158].


\bibitem{Kaloper:2002cs}
  N.~Kaloper, M.~Kleban, A.~Lawrence, S.~Shenker and L.~Susskind,
  JHEP {\bf 0211}, 037 (2002)
  [arXiv:hep-th/0209231].

\bibitem{Polchinskicosmic}
  J.~Polchinski,
  arXiv:hep-th/0412244.


\bibitem{Tye:2005wv}
  S.~H.~H.~Tye,
  AIP Conf.\ Proc.\  {\bf 743}, 410 (2005).


\bibitem{Polchinski:2007qc}
  J.~Polchinski,
  arXiv:0707.0888 [astro-ph].


\bibitem{KKLMMT}
  S.~Kachru, R.~Kallosh, A.~Linde, J.~M.~Maldacena, L.~McAllister and S.~P.~Trivedi,
  JCAP {\bf 0310}, 013 (2003)
  [arXiv:hep-th/0308055].



\bibitem{Sarangi:2002yt}
  S.~Sarangi and S.~H.~H.~Tye,
  Phys.\ Lett.\  B {\bf 536}, 185 (2002)
  [arXiv:hep-th/0204074].

\bibitem{Copeland:2004iv}
  E.~J.~Copeland, R.~C.~Myers and J.~Polchinski,
  Comptes Rendus Physique {\bf 5} (2004) 1021.



\bibitem{MoreCosmicStrings}
  L.~Leblond and S.~H.~H.~Tye,
  JHEP {\bf 0403}, 055 (2004)
  [arXiv:hep-th/0402072].

  N.~Barnaby, A.~Berndsen, J.~M.~Cline and H.~Stoica,
  JHEP {\bf 0506}, 075 (2005)
  [arXiv:hep-th/0412095].

  H.~Firouzjahi and S.~H.~Tye,
  JCAP {\bf 0503}, 009 (2005)
  [arXiv:hep-th/0501099].

  S.~H.~Tye, I.~Wasserman and M.~Wyman,
  Phys.\ Rev.\  D {\bf 71}, 103508 (2005)
  [Erratum-ibid.\  D {\bf 71}, 129906 (2005)]
  [arXiv:astro-ph/0503506].

  H.~Firouzjahi, L.~Leblond and S.~H.~Henry Tye,
  JHEP {\bf 0605}, 047 (2006)
  [arXiv:hep-th/0603161].

  L.~Leblond and M.~Wyman,
  arXiv:astro-ph/0701427.

\bibitem{DamourVilenkin}
  T.~Damour and A.~Vilenkin,
  Phys.\ Rev.\  D {\bf 64}, 064008 (2001)
  [arXiv:gr-qc/0104026].

  T.~Damour and A.~Vilenkin,
  Phys.\ Rev.\ Lett.\  {\bf 85}, 3761 (2000)
  [arXiv:gr-qc/0004075].


\bibitem{Wecht}
B. Wecht,
%
%
arXiv:0708.3984 [hep-th].


\bibitem{Svrcek:2006yi}
  P.~Svrcek and E.~Witten,
  JHEP {\bf 0606}, 051 (2006)
  [arXiv:hep-th/0605206].

\bibitem{CopelandEta}
  E.~J.~Copeland, A.~R.~Liddle, D.~H.~Lyth, E.~D.~Stewart and D.~Wands,
  Phys.\ Rev.\  D {\bf 49}, 6410 (1994)
  [arXiv:astro-ph/9401011].

\bibitem{Ganor}
  O.~J.~Ganor,
  Nucl.\ Phys.\ B {\bf 499}, 55 (1997) [arXiv:hep-th/9612077].

\bibitem{BHK}
  M.~Berg, M.~Haack and B.~Kors,
  Phys.\ Rev.\ D {\bf 71}, 026005 (2005)
  [arXiv:hep-th/0404087].

\bibitem{GM}
S.~B.~Giddings and A.~Maharana,
  Phys.\ Rev.\ D {\bf 73}, 126003 (2006)
  [arXiv:hep-th/0507158].


\bibitem{BDKMMM}
  D.~Baumann, A.~Dymarsky, I.~R.~Klebanov, J.~Maldacena, L.~McAllister and A.~Murugan,
  JHEP {\bf 0611}, 031 (2006)
  [arXiv:hep-th/0607050].



\bibitem{BDKMS}
  D.~Baumann, A.~Dymarsky, I.~R.~Klebanov, L.~McAllister and P.~J.~Steinhardt,
  arXiv:0705.3837 [hep-th].




\bibitem{BHKKahler}

K.~Becker, M.~Becker, M.~Haack and J.~Louis,
  JHEP {\bf 0206}, 060 (2002)
  [arXiv:hep-th/0204254].

M.~Berg, M.~Haack and B.~Kors,
  JHEP {\bf 0511}, 030 (2005)
  [arXiv:hep-th/0508043].

\bibitem{Kinney}
  S.~Dodelson, W.~H.~Kinney and E.~W.~Kolb,
  Phys.\ Rev.\  D {\bf 56}, 3207 (1997)
  [arXiv:astro-ph/9702166].

\bibitem{Neupane:2005nb}
  I.~P.~Neupane and D.~L.~Wiltshire,
  Phys.\ Rev.\  D {\bf 72}, 083509 (2005)
  [arXiv:hep-th/0504135].


\bibitem{FieldRange}
  D.~Baumann and L.~McAllister,
  arXiv:hep-th/0610285.


\bibitem{BLS}

M.~Becker, L.~Leblond and S.~E.~Shandera,
  arXiv:0709.1170 [hep-th].

T.~Kobayashi, S.~Mukohyama and S.~Kinoshita,
  arXiv:0708.4285 [hep-th].




\bibitem{Nflation}
  S.~Dimopoulos, S.~Kachru, J.~McGreevy and J.~G.~Wacker,
 arXiv:hep-th/0507205.


\bibitem{KL}
  R.~Kallosh and A.~Linde,
  JCAP {\bf 0704}, 017 (2007)
  [arXiv:0704.0647 [hep-th]].

\bibitem{Dvali:1998pa}
  G.~R.~Dvali and S.~H.~H.~Tye,
  Phys.\ Lett.\  B {\bf 450}, 72 (1999)
  [arXiv:hep-ph/9812483].


\bibitem{Alexander:2001ks}
  S.~H.~S.~Alexander,
  Phys.\ Rev.\  D {\bf 65}, 023507 (2002)
  [arXiv:hep-th/0105032].

\bibitem{Dvali:2001fw}
  G.~R.~Dvali, Q.~Shafi and S.~Solganik,
  arXiv:hep-th/0105203.


\bibitem{SomeAspects}
  G.~Shiu and S.~H.~H.~Tye,
  Phys.\ Lett.\  B {\bf 516}, 421 (2001)
  [arXiv:hep-th/0106274].

\bibitem{Burgess:2001fx}
  C.~P.~Burgess, M.~Majumdar, D.~Nolte, F.~Quevedo, G.~Rajesh and R.~J.~Zhang,
  JHEP {\bf 0107}, 047 (2001)
  [arXiv:hep-th/0105204].

\bibitem{Burgess:2001vr}
  C.~P.~Burgess, P.~Martineau, F.~Quevedo, G.~Rajesh and R.~J.~Zhang,
  JHEP {\bf 0203}, 052 (2002)
  [arXiv:hep-th/0111025].

\bibitem{Jones}
  N.~T.~Jones, H.~Stoica and S.~H.~H.~Tye,
  JHEP {\bf 0207}, 051 (2002)
  [arXiv:hep-th/0203163].

\bibitem{Oldmoduli}
P.~Binetruy and M.~K.~Gaillard,
  Phys.\ Rev.\  D {\bf 34}, 3069 (1986).

T.~Banks, M.~Berkooz, S.~H.~Shenker, G.~W.~Moore and P.~J.~Steinhardt,
  Phys.\ Rev.\  D {\bf 52}, 3548 (1995)
  [arXiv:hep-th/9503114].


\bibitem{BKQ}
  C.~P.~Burgess, R.~Kallosh and F.~Quevedo,
  JHEP {\bf 0310}, 056 (2003)
  [arXiv:hep-th/0309187].

\bibitem{VZ}
  G.~Villadoro and F.~Zwirner,
  Phys.\ Rev.\ Lett.\  {\bf 95}, 231602 (2005)
  [arXiv:hep-th/0508167].


\bibitem{KPV}
  S.~Kachru, J.~Pearson and H.~L.~Verlinde,
  JHEP {\bf 0206}, 021 (2002)
  [arXiv:hep-th/0112197].

\bibitem{DKS}
  A.~Dymarsky, I.~R.~Klebanov and N.~Seiberg,
  JHEP {\bf 0601}, 155 (2006)
  [arXiv:hep-th/0511254].


\bibitem{Fuplift}

A.~Saltman and E.~Silverstein,
  JHEP {\bf 0411}, 066 (2004)
  [arXiv:hep-th/0402135].

  K.~Becker, Y.~C.~Chung and G.~y.~Guo,
  arXiv:0706.2502 [hep-th].




\bibitem{ThreeSeven}

  C.~Herdeiro, S.~Hirano and R.~Kallosh,
  JHEP {\bf 0112}, 027 (2001)
  [arXiv:hep-th/0110271].

  K.~Dasgupta, C.~Herdeiro, S.~Hirano and R.~Kallosh,
  Phys.\ Rev.\  D {\bf 65}, 126002 (2002)
  [arXiv:hep-th/0203019].


  K.~Dasgupta, J.~P.~Hsu, R.~Kallosh, A.~Linde and M.~Zagermann,
  JHEP {\bf 0408}, 030 (2004)
  [arXiv:hep-th/0405247].


\bibitem{DeWolfeGiddings}
  O.~DeWolfe and S.~B.~Giddings,
  Phys.\ Rev.\  D {\bf 67}, 066008 (2003)
  [arXiv:hep-th/0208123].


\bibitem{ThreeSevenShift}
  J.~P.~Hsu, R.~Kallosh and S.~Prokushkin,
  JCAP {\bf 0312}, 009 (2003)
  [arXiv:hep-th/0311077].

  J.~P.~Hsu and R.~Kallosh,
  JHEP {\bf 0404}, 042 (2004)
  [arXiv:hep-th/0402047].

  S.~E.~Shandera,
  JCAP {\bf 0504}, 011 (2005)
  [arXiv:hep-th/0412077].

\bibitem{ThreeSevenMass}
  M.~Berg, M.~Haack and B.~Kors,
  arXiv:hep-th/0409282.

  L.~McAllister,
  JCAP {\bf 0602}, 010 (2006)
  [arXiv:hep-th/0502001].

\bibitem{AspinwallKallosh}
  P.~S.~Aspinwall and R.~Kallosh,
  JHEP {\bf 0510}, 001 (2005)
  [arXiv:hep-th/0506014].

\bibitem{Angles1}
  J.~Garcia-Bellido, R.~Rabadan and F.~Zamora,
  JHEP {\bf 0201}, 036 (2002)
  [arXiv:hep-th/0112147].

\bibitem{Angles2}
  R.~Blumenhagen, B.~Kors, D.~Lust and T.~Ott,
  Nucl.\ Phys.\  B {\bf 641}, 235 (2002)
  [arXiv:hep-th/0202124].


\bibitem{HardToCategorize}




  D.~Choudhury, D.~Ghoshal, D.~P.~Jatkar and S.~Panda,
  JCAP {\bf 0307}, 009 (2003)
  [arXiv:hep-th/0305104].

  L.~Pilo, A.~Riotto and A.~Zaffaroni,
  JHEP {\bf 0407}, 052 (2004)
  [arXiv:hep-th/0401004].





  D.~Cremades, F.~Quevedo and A.~Sinha,
  JHEP {\bf 0510}, 106 (2005)
  [arXiv:hep-th/0505252].

  A.~Avgoustidis, D.~Cremades and F.~Quevedo,
  arXiv:hep-th/0606031.


  N.~Barnaby, T.~Biswas and J.~M.~Cline,
  JHEP {\bf 0704}, 056 (2007)
  [arXiv:hep-th/0612230].


  B.~Dutta, J.~Kumar and L.~Leblond,
  arXiv:hep-th/0703278.






\bibitem{KS}
  I.~R.~Klebanov and M.~J.~Strassler,
  JHEP {\bf 0008}, 052 (2000)
  [arXiv:hep-th/0007191].


\bibitem{FurtherWarped}
  S.~Shandera, B.~Shlaer, H.~Stoica and S.~H.~H.~Tye,
  JCAP {\bf 0402}, 013 (2004)
  [arXiv:hep-th/0311207].

  H.~Firouzjahi and S.~H.~H.~Tye,
  Phys.\ Lett.\  B {\bf 584}, 147 (2004)
  [arXiv:hep-th/0312020].

  C.~P.~Burgess, J.~M.~Cline, H.~Stoica and F.~Quevedo,
  JHEP {\bf 0409}, 033 (2004)
  [arXiv:hep-th/0403119].

  S.~E.~Shandera and S.~H.~Tye,
  JCAP {\bf 0605}, 007 (2006)
  [arXiv:hep-th/0601099].

  X.~Chen, S.~Sarangi, S.~H.~Henry Tye and J.~Xu,
  JCAP {\bf 0611}, 015 (2006)
  [arXiv:hep-th/0608082].

  L.~Leblond and S.~Shandera,
  JCAP {\bf 0701}, 009 (2007)
  [arXiv:hep-th/0610321].

 G.~Hailu and S.~H.~Tye,
  JHEP {\bf 0708}, 009 (2007)
  [arXiv:hep-th/0611353].

\bibitem{WarpedReheating}



  N.~Barnaby and J.~M.~Cline,
  Int.\ J.\ Mod.\ Phys.\  A {\bf 19}, 5455 (2004)
  [arXiv:hep-th/0410030].

  N.~Barnaby, C.~P.~Burgess and J.~M.~Cline,
  JCAP {\bf 0504}, 007 (2005)
  [arXiv:hep-th/0412040].


  L.~Kofman and P.~Yi,
  Phys.\ Rev.\  D {\bf 72}, 106001 (2005)
  [arXiv:hep-th/0507257].

  A.~R.~Frey, A.~Mazumdar and R.~Myers,
  Phys.\ Rev.\  D {\bf 73}, 026003 (2006)
  [arXiv:hep-th/0508139].

  D.~Chialva, G.~Shiu and B.~Underwood,
  JHEP {\bf 0601}, 014 (2006)
  [arXiv:hep-th/0508229].


  X.~Chen and S.~H.~Tye,
  JCAP {\bf 0606}, 011 (2006)
  [arXiv:hep-th/0602136].


\bibitem{GiantInflaton}
  O.~DeWolfe, S.~Kachru and H.~L.~Verlinde,
  JHEP {\bf 0405}, 017 (2004)
  [arXiv:hep-th/0403123].

\bibitem{Sandip}
  N.~Iizuka and S.~P.~Trivedi,
  Phys.\ Rev.\ D {\bf 70}, 043519 (2004)  [arXiv:hep-th/0403203];


\bibitem{Cline:2005ty}
  J.~M.~Cline and H.~Stoica,
  Phys.\ Rev.\  D {\bf 72}, 126004 (2005)
  [arXiv:hep-th/0508029].




\bibitem{Burgess}
  C.~P.~Burgess, J.~M.~Cline, K.~Dasgupta and H.~Firouzjahi,
  JHEP {\bf 0703}, 027 (2007)
  [arXiv:hep-th/0610320].




\bibitem{Ouyang}
  P.~Ouyang,
  Nucl.\ Phys.\ B {\bf 699}, 207 (2004)
  [arXiv:hep-th/0311084].


\bibitem{KrausePajer}
  A.~Krause and E.~Pajer,
  arXiv:0705.4682v2 [hep-th].

\bibitem{Kuperstein}
  S.~Kuperstein,
  JHEP {\bf 0503}, 014 (2005) [arXiv:hep-th/0411097].

\bibitem{Spalinski}

  M.~Spalinski,
  JCAP {\bf 0704}, 018 (2007)
  [arXiv:hep-th/0702118].

  M.~Spalinski,
  JCAP {\bf 0705}, 017 (2007)
  [arXiv:hep-th/0702196].

M.~Spalinski,
  Phys.\ Lett.\  B {\bf 650}, 313 (2007)
  [arXiv:hep-th/0703248].

\bibitem{ChenDBI}

  X.~Chen,
  Phys.\ Rev.\  D {\bf 71}, 026008 (2005)
  [arXiv:hep-th/0406198].

  X.~Chen,
  Phys.\ Rev.\  D {\bf 71}, 063506 (2005)
  [arXiv:hep-th/0408084].

  X.~Chen,
  JHEP {\bf 0508}, 045 (2005)
  [arXiv:hep-th/0501184].

 X.~Chen,
  Phys.\ Rev.\  D {\bf 72}, 123518 (2005)
  [arXiv:astro-ph/0507053].



\bibitem{Bean:2007hc}
  R.~Bean, S.~E.~Shandera, S.~H.~Henry Tye and J.~Xu,
  JCAP {\bf 0705}, 004 (2007)
  [arXiv:hep-th/0702107].

\bibitem{Lidsey:2007gq}
  J.~E.~Lidsey and I.~Huston,
  JCAP {\bf 0707} (2007) 002
  [arXiv:0705.0240 [hep-th]].

\bibitem{Peiris:2007gz}
  H.~V.~Peiris, D.~Baumann, B.~Friedman and A.~Cooray,
  arXiv:0706.1240 [astro-ph].

\bibitem{Spinflation}
  D.~A.~Easson, R.~Gregory, D.~F.~Mota, G.~Tasinato and I.~Zavala,
  arXiv:0709.2666 [hep-th].

\bibitem{ThomasWard}
  S.~Thomas and J.~Ward,
  Phys.\ Rev.\  D {\bf 76}, 023509 (2007)
  [arXiv:hep-th/0702229].

  M.~x.~Huang, G.~Shiu and B.~Underwood,
  arXiv:0709.3299 [hep-th].




\bibitem{ShiuUnderwood}
  S.~Kecskemeti, J.~Maiden, G.~Shiu and B.~Underwood,
  JHEP {\bf 0609}, 076 (2006)
  [arXiv:hep-th/0605189].


  G.~Shiu and B.~Underwood,
  Phys.\ Rev.\ Lett.\  {\bf 98}, 051301 (2007)
  [arXiv:hep-th/0610151].

L.~Lorenz, J.~Martin and C.~Ringeval,
  arXiv:0709.3758 [hep-th].


  R.~Bean, X.~Chen, H.~V.~Peiris and J.~Xu,
  arXiv:0710.1812 [hep-th].

  F.~Gmeiner and C.~D.~White,
  arXiv:0710.2009 [hep-th].






\bibitem{Brown:2007ce}
  A.~R.~Brown, S.~Sarangi, B.~Shlaer and A.~Weltman,
  arXiv:0706.0485 [hep-fth].


\bibitem{Sarangi:2007mj}
  S.~Sarangi,
  arXiv:0710.0421 [hep-th].


\bibitem{DBIassist}
H.~Singh,
  Nucl.\ Phys.\  B {\bf 734}, 169 (2006)
  [arXiv:hep-th/0508101].

H.~Singh,
  arXiv:hep-th/0608032.

K.~L.~Panigrahi and H.~Singh,
  arXiv:0708.1679 [hep-th].



\bibitem{SingleM5}
  E.~I.~Buchbinder,
  Nucl.\ Phys.\  B {\bf 711}, 314 (2005)
  [arXiv:hep-th/0411062].

\bibitem{AssistedM5}
  K.~Becker, M.~Becker and A.~Krause,
  Nucl.\ Phys.\  B {\bf 715}, 349 (2005)
  [arXiv:hep-th/0501130].

\bibitem{Axel}
  A.~Krause,
  arXiv:0708.4414 [hep-th].

\bibitem{Kahler}
  J.~P.~Conlon and F.~Quevedo,
  JHEP {\bf 0601}, 146 (2006)
  [arXiv:hep-th/0509012].


\bibitem{BBCQ}
  V.~Balasubramanian, P.~Berglund, J.~P.~Conlon and F.~Quevedo,
  JHEP {\bf 0503}, 007 (2005)
  [arXiv:hep-th/0502058].

\bibitem{Roulette}
  J.~R.~Bond, L.~Kofman, S.~Prokushkin and P.~M.~Vaudrevange,
  arXiv:hep-th/0612197.


\bibitem{LoopsOne}
  M.~Berg, M.~Haack and E.~Pajer,
  arXiv:0704.0737 [hep-th].

\bibitem{LoopsTwo}
  M.~Cicoli, J.~P.~Conlon and F.~Quevedo,
  arXiv:0708.1873 [hep-th].

\bibitem{DGKT}
  O.~DeWolfe, A.~Giryavets, S.~Kachru and W.~Taylor,
  JHEP {\bf 0507}, 066 (2005)
  [arXiv:hep-th/0505160].

\bibitem{IW}
  M.~Ihl and T.~Wrase,
  JHEP {\bf 0607}, 027 (2006)
  [arXiv:hep-th/0604087].

\bibitem{Hertzberg}
  M.~P.~Hertzberg, M.~Tegmark, S.~Kachru, J.~Shelton and O.~Ozcan,
  arXiv:0709.0002 [astro-ph].


\bibitem{Natural}
  K.~Freese, J.~A.~Frieman and A.~V.~Olinto,
  Phys.\ Rev.\ Lett.\  {\bf 65}, 3233 (1990).

\bibitem{Gorbatov}
  T.~Banks, M.~Dine, P.~J.~Fox and E.~Gorbatov,
  JCAP {\bf 0306}, 001 (2003)
  [arXiv:hep-th/0303252].


\bibitem{CompletingNatural}
  J.~E.~Kim, H.~P.~Nilles and M.~Peloso,
  JCAP {\bf 0501}, 005 (2005)
  [arXiv:hep-ph/0409138].

\bibitem{Racetrack}
  J.~J.~Blanco-Pillado {\it et al.},
  JHEP {\bf 0411}, 063 (2004)
  [arXiv:hep-th/0406230].


\bibitem{BetterRacetrack}
  J.~J.~Blanco-Pillado {\it et al.},
  JHEP {\bf 0609}, 002 (2006)
  [arXiv:hep-th/0603129].


\bibitem{DDF}
  F.~Denef, M.~R.~Douglas and B.~Florea,
  JHEP {\bf 0406}, 034 (2004)
  [arXiv:hep-th/0404257].

\bibitem{Amanda}
  B.~Greene and A.~Weltman,
  JHEP {\bf 0603}, 035 (2006)
  [arXiv:hep-th/0512135].


\bibitem{RMTNflation}
  R.~Easther and L.~McAllister,
  JCAP {\bf 0605}, 018 (2006)
  [arXiv:hep-th/0512102].

\bibitem{KimLiddleOne}
  S.~A.~Kim and A.~R.~Liddle,
  Phys.\ Rev.\  D {\bf 74}, 023513 (2006)
  [arXiv:astro-ph/0605604].


\bibitem{Piao}
  Y.~S.~Piao,
  Phys.\ Rev.\  D {\bf 74}, 047302 (2006)
  [arXiv:gr-qc/0606034].

\bibitem{KimLiddleTwo}
  S.~A.~Kim and A.~R.~Liddle,
  Phys.\ Rev.\  D {\bf 74}, 063522 (2006)
  [arXiv:astro-ph/0608186].


\bibitem{Battefeld}
  D.~Battefeld and T.~Battefeld,
  JCAP {\bf 0705}, 012 (2007)
  [arXiv:hep-th/0703012].

\bibitem{KimLiddleThree}
  S.~A.~Kim and A.~R.~Liddle,
  arXiv:0707.1982 [astro-ph].


\bibitem{RenataReview}
  R.~Kallosh,
  arXiv:hep-th/0702059.


\bibitem{Green:2007gs}
  D.~Green,
  arXiv:0707.3832 [hep-th].




\bibitem{NewOldOne}
  G.~Dvali and S.~Kachru,
  arXiv:hep-th/0309095.

\bibitem{NewOldTwo}
  G.~Dvali and S.~Kachru,
  arXiv:hep-ph/0310244.

\bibitem{LockedOne}
  R.~Easther, J.~Khoury and K.~Schalm,
  JCAP {\bf 0406}, 006 (2004)
  [arXiv:hep-th/0402218].

\bibitem{LockedTwo}
  E.~J.~Copeland and A.~Rajantie,
  JCAP {\bf 0502}, 008 (2005)
  [arXiv:astro-ph/0501668].

\bibitem{Multiple}
  C.~P.~Burgess, R.~Easther, A.~Mazumdar, D.~F.~Mota and T.~Multamaki,
  JHEP {\bf 0505}, 067 (2005)
  [arXiv:hep-th/0501125].


\bibitem{ChainOne}
  K.~Freese and D.~Spolyar,
  JCAP {\bf 0507}, 007 (2005)
  [arXiv:hep-ph/0412145].

\bibitem{ChainTwo}
  B.~Feldstein and B.~Tweedie,
  JCAP {\bf 0704}, 020 (2007)
  [arXiv:hep-ph/0611286].

\bibitem{ChainThree}
  K.~Freese, J.~T.~Liu and D.~Spolyar,
  arXiv:hep-th/0612056.

\bibitem{ChainFour}
  Q.~G.~Huang,
  arXiv:0704.2835 [hep-th].


\bibitem{CyclicReviewOne}
  P.~J.~Steinhardt and N.~Turok,
  Nucl.\ Phys.\ Proc.\ Suppl.\  {\bf 124}, 38 (2003)
  [arXiv:astro-ph/0204479].


\bibitem{CyclicReviewTwo}
  P.~J.~Steinhardt and N.~Turok,
  New Astron.\ Rev.\  {\bf 49}, 43 (2005)
  [arXiv:astro-ph/0404480].

\bibitem{AndreiEkpyrotic}
  A.~Linde,
  arXiv:hep-th/0205259.

\bibitem{Mirage}
  A.~Kehagias and E.~Kiritsis,
  JHEP {\bf 9911}, 022 (1999)
  [arXiv:hep-th/9910174].

\bibitem{ScottReview}
  T.~Battefeld and S.~Watson,
  Rev.\ Mod.\ Phys.\  {\bf 78}, 435 (2006)
  [arXiv:hep-th/0510022].

\bibitem{PBB}
  M.~Gasperini and G.~Veneziano,
  arXiv:hep-th/0703055.








\bibitem{Traschen:1990sw}
  J.~H.~Traschen and R.~H.~Brandenberger,
  Phys.\ Rev.\  D {\bf 42} (1990) 2491.


\bibitem{KLSpreheating}

  L.~Kofman, A.~D.~Linde and A.~A.~Starobinsky,
  Phys.\ Rev.\ Lett.\  {\bf 73} (1994) 3195
  [arXiv:hep-th/9405187].


  L.~Kofman, A.~D.~Linde and A.~A.~Starobinsky,
  Phys.\ Rev.\ Lett.\  {\bf 76} (1996) 1011
  [arXiv:hep-th/9510119].


  L.~Kofman, A.~D.~Linde and A.~A.~Starobinsky,
  arXiv:hep-ph/9608341.


  L.~Kofman, A.~D.~Linde and A.~A.~Starobinsky,
  Phys.\ Rev.\  D {\bf 56} (1997) 3258
  [arXiv:hep-ph/9704452].



\bibitem{KLLMMS}
  L.~Kofman, A.~Linde, X.~Liu, A.~Maloney, L.~McAllister and E.~Silverstein,
  JHEP {\bf 0405}, 030 (2004)
  [arXiv:hep-th/0403001].


\bibitem{Watson:2004aq}
  S.~Watson,
  Phys.\ Rev.\  D {\bf 70}, 066005 (2004)
  [arXiv:hep-th/0404177].

\bibitem{Greene:2007sa}
  B.~Greene, S.~Judes, J.~Levin, S.~Watson and A.~Weltman,
  JHEP {\bf 0707}, 060 (2007)
  [arXiv:hep-th/0702220].

\bibitem{DineSymmetries}


  M.~Dine, Y.~Nir and Y.~Shadmi,
  Phys.\ Lett.\  B {\bf 438}, 61 (1998)
  [arXiv:hep-th/9806124].

 M.~Dine,
  Phys.\ Lett.\  B {\bf 482}, 213 (2000)
  [arXiv:hep-th/0002047].

\bibitem{McAllister:2004gd}
  L.~McAllister and I.~Mitra,
  JHEP {\bf 0502}, 019 (2005)
  [arXiv:hep-th/0408085].


\bibitem{Brandenberger:1988aj}
  R.~H.~Brandenberger and C.~Vafa,
  Nucl.\ Phys.\  B {\bf 316}, 391 (1989).


\bibitem{Senreview}
  A.~Sen,
  Phys.\ Scripta {\bf T117}, 70 (2005)
  [arXiv:hep-th/0312153].



\bibitem{Freedman:2005wx}
  D.~Z.~Freedman, M.~Headrick and A.~Lawrence,
  Phys.\ Rev.\  D {\bf 73}, 066015 (2006)
  [arXiv:hep-th/0510126].

\bibitem{Myers:1987fv}
  R.~C.~Myers,
  Phys.\ Lett.\  B {\bf 199} (1987) 371.



\bibitem{otheroldSC}
  J.~Polchinski,
  Nucl.\ Phys.\  B {\bf 324}, 123 (1989).

  I.~Antoniadis, C.~Bachas, J.~R.~Ellis and D.~V.~Nanopoulos,
  Nucl.\ Phys.\  B {\bf 328}, 117 (1989).

  I.~Antoniadis, C.~Bachas, J.~R.~Ellis and D.~V.~Nanopoulos,
  Phys.\ Lett.\  B {\bf 257}, 278 (1991).

  A.~R.~Cooper, L.~Susskind and L.~Thorlacius,
  Nucl.\ Phys.\  B {\bf 363}, 132 (1991).

  S.~Mukherji,
  Mod.\ Phys.\ Lett.\  A {\bf 7}, 1361 (1992)
  [arXiv:hep-th/9203010].

  A.~A.~Tseytlin,
  Int.\ J.\ Mod.\ Phys.\  D {\bf 1}, 223 (1992)
  [arXiv:hep-th/9203033].

  C.~Schmidhuber and A.~A.~Tseytlin,
  Nucl.\ Phys.\  B {\bf 426}, 187 (1994)
  [arXiv:hep-th/9404180].


\bibitem{Creminelli:2007aq}
  P.~Creminelli and L.~Senatore,
  arXiv:hep-th/0702165.

\bibitem{Itzhaki}
  N.~Itzhaki and E.~D.~Kovetz,
  arXiv:0708.2798 [hep-th].


\bibitem{DineSeiberg}

  M.~Dine and N.~Seiberg,
  Phys.\ Lett.\  B {\bf 162}, 299 (1985).


\bibitem{laterlenny}
  N.~Goheer, M.~Kleban and L.~Susskind,
  JHEP {\bf 0307}, 056 (2003)
  [arXiv:hep-th/0212209].

\bibitem{Westphal:2007xd}
  A.~Westphal,
  arXiv:0705.1557 [hep-th].



\bibitem{Saltman:2004jh}
  A.~Saltman and E.~Silverstein,
  JHEP {\bf 0601}, 139 (2006)
  [arXiv:hep-th/0411271].

\bibitem{StromingerD}
A.~Strominger,
Phys.\ Rev.\ D {\bf 24}, 3082 (1981).

\bibitem{split}
  N.~Arkani-Hamed and S.~Dimopoulos,
  JHEP {\bf 0506}, 073 (2005)
  [arXiv:hep-th/0405159].



\bibitem{Conlon:2004ds}
  J.~P.~Conlon and F.~Quevedo,
  JHEP {\bf 0410}, 039 (2004)
  [arXiv:hep-th/0409215].

\bibitem{Denef:2005mm}
  F.~Denef, M.~R.~Douglas, B.~Florea, A.~Grassi and S.~Kachru,
  Adv.\ Theor.\ Math.\ Phys.\  {\bf 9}, 861 (2005)
  [arXiv:hep-th/0503124].

\bibitem{ESdS}
E.~Silverstein,
  ``Simple de Sitter Solutions,''
  arXiv:0712.1196 [hep-th].






\bibitem{Weinberg:2000yb}
  S.~Weinberg,
  arXiv:astro-ph/0005265.


\bibitem{Khoury:2003rn}
D.~F.~Mota and J.~D.~Barrow,
  Phys.\ Lett.\  B {\bf 581}, 141 (2004)
  [arXiv:astro-ph/0306047].

D.~F.~Mota and J.~D.~Barrow,
  Mon.\ Not.\ Roy.\ Astron.\ Soc.\  {\bf 349}, 291 (2004)
  [arXiv:astro-ph/0309273].

J.~Khoury and A.~Weltman,
  Phys.\ Rev.\  D {\bf 69}, 044026 (2004)
  [arXiv:astro-ph/0309411].

\bibitem{Freivogel:2005vv}
  B.~Freivogel, M.~Kleban, M.~Rodriguez Martinez and L.~Susskind,
  JHEP {\bf 0603}, 039 (2006)
  [arXiv:hep-th/0505232].


\end{thebibliography}
\end{document}